\documentclass[dvips]{arxstspdf}
\usepackage{graphicx}
\usepackage{flushend}
\usepackage{stfloats}

\volume{25}
\issue{3}
\pubyear{2010}
\firstpage{289}
\lastpage{310}
\doi{10.1214/10-STS330}

\makeatletter
\newcommand{\cal}{\mathcal}
\renewcommand{\epsilon}{\varepsilon}
\makeatother

\begin{document}
\begin{frontmatter}

\title{To Explain or to Predict?}
\runtitle{To Explain or to Predict?}

\begin{aug}
\author{\fnms{Galit} \snm{Shmueli}\corref{}\ead[label=e1]{gshmueli@umd.edu}}
\runauthor{G. Shmueli}

\affiliation{University of Maryland}

\address{Galit Shmueli is Associate Professor of Statistics,
Department of Decision, Operations and Information Technologies,
Robert H. Smith School of Business,
University of Maryland, College Park, Maryland~20742, USA
(\printead{e1}).}

\end{aug}

% ABSTRACT
%
\begin{abstract}
Statistical modeling is a powerful tool for developing and testing
theories by way of causal explanation, prediction, and description. In
many disciplines there is near-exclusive use of statistical modeling
for causal explanation and the assumption that models with high
explanatory power are inherently of high predictive power. Conflation
between explanation and prediction is common, yet the distinction must
be understood for progressing scientific knowledge. While this
distinction has been recognized in the philosophy of science, the
statistical literature lacks a thorough discussion of the many
differences that arise in the process of modeling for an explanatory
versus a predictive goal. %The fact that the ``best explanatory model''
%will almost always differ substantially from the ``best predictive
%model'' %requires an explanation of the sources of the difference
%between explanation and prediction and their implications to %each
%step in the modeling process.
The purpose of this article is to clarify the distinction between
explanatory and predictive modeling, to discuss its sources, and to
reveal the practical implications of the distinction to each step in
the modeling process.
%Although a vast literature exists on statistical modeling, on good
%practices, and on abuses of statistical models, the %literature lacks
%the discussion of a key component: the distinction between modeling
%for explanatory purposes and
%modeling for predictive purposes.
%This omission exacts considerable cost in terms of advancing
%scientific research in many fields. The goal of this paper %is to
%highlight these two uses of empirical modeling in scientific research
%and to examine the differences that arise %in each of the two
%statistical modeling paths. % distinction between the two modeling
%paths.
% and to emphasize the necessity of both explanatory and predictive
%modeling in scientific research.
%This distinction, which is often blurred in the statistics literature,
%has permeated into many fields of application, %hindering the
%potential usefulness of statistical modeling for scientific
%advancements.
\end{abstract}

% KEYWORDS
%

\end{frontmatter}

%s1 ###
\section{Introduction}
\label{sec-intro}
Looking at how statistical models are used in different scientific
disciplines for the purpose of theory building and testing, one finds a
range of perceptions regarding the relationship between causal
explanation and empirical prediction. In many scientific fields such as
economics, psychology, education, and environmental science,
statistical models are used almost exclusively for causal explanation,
and models that possess high explanatory power are often assumed to
inherently possess predictive power. In fields such as natural language
processing and bioinformatics, the focus is on empirical prediction
with only a slight and indirect relation to causal explanation. And yet
in other research fields, such as epidemiology, the emphasis on causal
explanation versus empirical prediction is more mixed. Statistical
modeling for description, where the purpose is to capture the data
structure parsimoniously, and which is the most commonly developed
within the field of statistics, is not commonly used for theory
building and testing in other disciplines. Hence, in this article I
focus on the use of statistical modeling for causal explanation and for
prediction. My main premise is that the two are often conflated, yet
the causal versus predictive distinction has a large impact on each
step of the statistical modeling process and on its consequences.
Although not explicitly stated in the statistics methodology
literature, applied statisticians instinctively sense that predicting
and explaining are different. This article aims to fill a critical
void: to tackle the distinction between explanatory modeling and
predictive modeling.

%f1 ###
\begin{figure*}[b]

\includegraphics{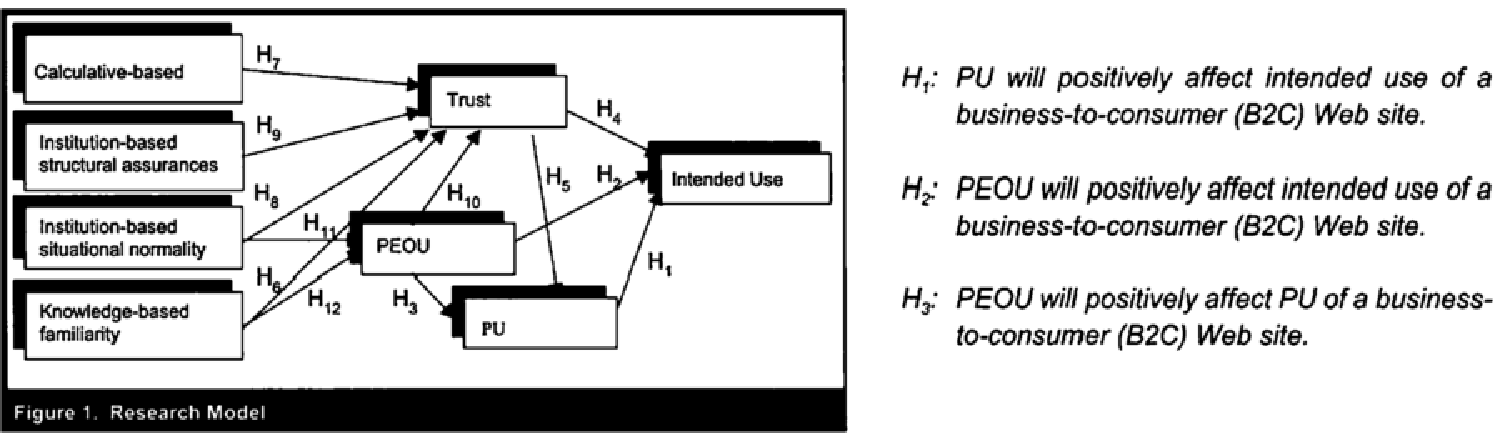}

\caption{Causal diagram \textup{(left)} and partial list of stated hypotheses
\textup{(right)} from Gefen, Karahanna and Straub (\protect\citeyear{Gefen2003}).}
\label{fig:causal-diagram}
\end{figure*}

Clearing the current ambiguity between the two is critical not only for
proper statistical modeling, but more importantly, for proper
scientific usage. Both explanation and prediction are necessary for
generating and testing theories, yet each plays a different role in
doing so.
%Hence, their current conflation has likely been hindering scientific
%advances in various fields.
The lack of a clear distinction within statistics has created a lack of
understanding in many disciplines of the difference between building
sound explanatory models versus creating powerful predictive models, as
well as confusing explanatory power with predictive power. The
implications of this omission and the lack of clear guidelines on how
to model for explanatory versus predictive goals are considerable for
both scientific research and practice
%. Many fields that use statistical models to build and advance
%scientific theory suffer from inadequate modeling, often %stemming
%from the conflation of explanatory and predictive goals and modeling.
%The indiscrimination between predicting %and explaining, and their
%respective roles,
and have also contributed to the gap between academia and practice.

%This dichotomy is pronounced in finance, for example, where practice
%is concerned with prediction whereas academic %research is focused on
%explanation.

I start by defining what I term \textit{explaining} and \textit
{predicting}. These definitions are chosen to reflect the distinct
scientific goals that they are aimed at: causal explanation and
empirical prediction, respectively. \textit{Explanatory modeling} and
\textit{predictive modeling} reflect the process of using data and
statistical (or data mining) methods for explaining or predicting,
respectively. The term \textit{modeling} is intentionally chosen over
\textit{models} to highlight the entire process involved, from goal
definition, study design, and data collection to scientific use.

%s1.1 ###
\subsection{Explanatory Modeling}
\label{subsec-explain}
In many scientific fields, and especially the social sciences,
statistical methods are used nearly exclusively for testing causal
theory. Given a causal theoretical model, statistical models are
applied to data in order to test causal hypotheses. In such models, a
set of underlying factors that are measured by variables $X$ are
assumed to cause an underlying effect, measured by variable $Y$. Based
on collaborative work with social scientists and economists, on an
examination of some of their literature, and on conversations with a
diverse group of researchers, I conjecture that, whether statisticians
like it or not, the type of statistical models used for testing causal
hypotheses in the social sciences are almost always association-based
models applied to observational data. Regression models are the most
common example. The justification for this practice is that the theory
itself provides the causality. In other words, the role of the theory
is very strong and the reliance on data and statistical modeling are
strictly through the lens of the theoretical model. The theory--data
relationship varies in different fields. While the social sciences are
very theory-heavy, in areas such as bioinformatics and natural language
processing the emphasis on a causal theory is much weaker.
Hence, given this reality, I define \textit{explaining} as causal
explanation and \textit{explanatory modeling} as the use of statistical
models for testing causal explanations.

To illustrate how explanatory modeling is typically done, I describe
the structure of a typical article in a highly regarded journal in the
field of Information Systems (IS). Researchers in the field of IS
usually have training in economics and/or the behavioral sciences. The
structure of articles reflects the way empirical research is conducted
in IS and related fields.

The example used is an article by Gefen, Karahanna and Straub (\citeyear{Gefen2003}), which studies
technology acceptance. The article starts with a presentation of the
prevailing relevant theory(ies):

\begin{quote}

Online purchase intensions should be explained in part by the
technology acceptance model (TAM). This theoretical model is at present
a preeminent theory of technology acceptance in IS.

\end{quote}

\noindent
The authors then proceed to state multiple causal hypotheses (denoted
$H_1, H_2,\ldots $ in Figure~\ref{fig:causal-diagram}, right panel),
justifying the merits for each hypothesis and grounding it in theory.
The research hypotheses are given in terms of theoretical \textit
{constructs} rather than measurable variables. Unlike measurable
variables, constructs are abstractions that
``describe a phenomenon of theoretical interest'' (Edwards and\break Bagozzi, \citeyear{Edwards2000})
and can be observable or unobservable. Examples of constructs in this
article are trust, perceived usefulness (PU), and perceived ease of use
(PEOU). Examples of constructs used in other fields include anger,
poverty, well-being, and odor. The hypotheses section will often
include a causal diagram illustrating the hypothesized causal
relationship between the constructs (see Figure~\ref{fig:causal-diagram},
left panel).
The next step is \textit{construct operationalization}, where a bridge
is built between theoretical constructs and observable measurements,
using previous literature and theoretical justification. Only after the
theoretical component is completed, and measurements are justified and
defined, do researchers proceed to the next step where data and
statistical modeling are introduced alongside the statistical
hypotheses, which are operationalized from the research hypotheses.
Statistical inference will lead to ``statistical conclusions'' in terms
of effect sizes and statistical significance in relation to the causal
hypotheses. Finally, the statistical conclusions are converted into
research conclusions, often accompanied by policy recommendations.

In summary, \textit{explanatory modeling} refers here to the
application of statistical models to data for testing causal hypotheses
about theoretical constructs. Whereas ``proper'' statistical methodology
for testing causality exists, such as designed experiments or
specialized causal inference methods for observational data [e.g.,
causal diagrams (Pearl, \citeyear{Pearl1995}), discovery algorithms
(Spirtes, Glymour and Scheines, \citeyear{Spirtes2000}), probability trees
(Shafer, \citeyear{Shafer1996}), and propensity
scores (Rosenbaum and Rubin, \citeyear{Rosenbaum1983}; Rubin, \citeyear{Rubin1997})],
 in practice association-based
statistical models, applied to observational data, are most commonly
used for that purpose.

%s1.2 ###
\subsection{Predictive Modeling}
\label{subsec-predict}
I define \textit{predictive modeling} as the process of applying a
statistical model or data mining algorithm to data for the purpose of
predicting new or future observations. In particular, I focus on
nonstochastic prediction (Geisser, \citeyear{Geisser1993}, page 31), where the goal
is to predict the output value ($Y$) for new observations given their
input values ($X$). This definition also includes temporal forecasting,
where observations until time $t$ (the input) are used to forecast
future values at time $t+k, k>0$ (the output). \textit{Predictions}
include point or interval predictions, prediction regions, predictive
distributions, or rankings of new observations. \textit{Predictive
model} is any method that produces predictions, regardless of its
underlying approach: Bayesian or frequentist, parametric or
nonparametric, data mining algorithm or statistical model, etc.

%Predictive modeling differs from explanatory modeling in two major
%ways: it is not based on causality, and it focuses %on measurable
%input and output variables ($X$ and $Y$) rather than on some
%underlying theoretical constructs.

%In examining the methodological literature of various fields, it
%appears that predictive modeling is nearly absent from %most social
%science fields and economics; it is somewhat present in engineering
%and medicine; and is very popular in %areas such as machine learning
%and artificial intelligence.

%s1.3 ###
\subsection{Descriptive Modeling}
\label{subsec-describe}
Although not the focus of this article, a third type of modeling, which
is the most commonly used and developed by statisticians, is
descriptive modeling. This type of modeling is aimed at summarizing or
representing the data structure in a compact manner. Unlike explanatory
modeling, in descriptive modeling the reliance on an underlying causal
theory is absent or incorporated in a less formal way. Also, the focus
is at the measurable level rather than at the construct level. Unlike
predictive modeling, descriptive modeling is not aimed at prediction.
%%Creating summary statistics and graphical visualizations as a final
%%product are examples.
Fitting a regression model can be descriptive if it is used for
capturing the association between the dependent and independent
variables rather than for causal inference or for prediction. We
mention this type of modeling to avoid confusion with
causal-explanatory and predictive modeling, and also to highlight the
different approaches of statisticians and nonstatisticians.

%s1.4 ###
\subsection{The Scientific Value of Predictive Modeling}

Although explanatory modeling is commonly used for theory building and
testing, predictive modeling is nearly absent in many scientific fields
as a tool for developing theory. One possible reason is the statistical
training of nonstatistician researchers. A look at many introductory
statistics textbooks reveals very little in the way of prediction.
Another reason is that prediction is often considered unscientific.
Berk (\citeyear{Berk2008}) wrote, ``In the social sciences, for example, one
either did causal modeling econometric style or largely gave up
quantitative work.'' From conversations with colleagues in various
disciplines it appears that predictive modeling is often valued for its
applied utility, yet is discarded for scientific purposes such as
theory building or testing. Shmueli and Koppius (\citeyear{ShmueliKoppius2010}) illustrated the
lack of predictive modeling in the field of IS. Searching the 1072
papers published in the two top-rated journals \textit{Information
Systems Research} and \textit{MIS Quarterly} between 1990 and 2006,
they found only 52 empirical papers with predictive claims, of which
only seven carried out proper predictive modeling or testing.

Even among academic statisticians, there appears to be a divide between
those who value prediction as the main purpose of statistical modeling
and those who see it as unacademic. Examples of statisticians who
emphasize predictive methodology include\break Akaike (``The predictive
point of view is a prototypical point of view to explain the basic
activity of statistical analysis'' in Findley and Parzen, \citeyear{Findley1998}), Deming
(``The only useful function of a statistician is to make predictions''
in Wallis, \citeyear{Wallis1980}), Geisser (``The prediction of observables or
potential observables is of much greater relevance than the estimate of
what are often artificial constructs-parameters,'' Geisser, \citeyear{Geisser1975}),
Aitchison and Dunsmore (``prediction analysis\ldots\  is surely at
the heart of many statistical applications,'' Aitchison and Dunsmore, \citeyear{Aitchison1975a})  and
Friedman (``One of the most common and important uses for data is
prediction,'' Friedman, \citeyear{Friedman1997}). Examples of those who see it as
unacademic are Kendall and Stuart (``The Science of Statistics
deals with the properties of populations. In considering a population
of men we are not interested, statistically speaking, in whether some
particular individual has brown eyes or is a forger, but rather in how
many of the individuals have brown eyes or are
forgers,'' Kendall and Stuart, \citeyear{Kendall1977}) and more recently Parzen (``The two
goals in analyzing data\ldots\  I prefer to describe as ``management'' and
``science.'' Management seeks profit\ldots\  Science seeks
truth,'' Parzen, \citeyear{Parzen2001}). In economics there is a similar disagreement
regarding ``whether prediction per se is a legitimate objective of
economic science, and also whether observed data should be used only to
shed light on existing theories or also for the purpose of hypothesis
seeking in order to develop new theories'' (Feelders, \citeyear{Feelders2002}).

Before proceeding with the discrimination between explanatory and
predictive modeling, it is important to establish prediction as a
necessary scientific endeavor beyond utility, for the purpose of
developing and testing theories.
Predictive modeling and predictive testing serve several necessary
scientific functions:
\begin{enumerate}[6.]
\item Newly available large and rich datasets often contain complex
relationships and patterns that are hard to hypothesize, especially
given theories that exclude newly measurable concepts. Using predictive
modeling in such contexts can help uncover potential new causal
mechanisms and lead to the generation of new hypotheses. See, for
example, the discussion between Gurbaxani and\break Mendelson (\citeyear{Gurbaxani1990,Gurbaxani1994}) and
Collopy, Adya  and Armstrong (\citeyear{Collopy1994}).
\item The development of new theory often goes hand in hand with the
development of new measures (Van Maanen, Sorensen and Mitchell, \citeyear{Van2007}). Predictive modeling can be
used to discover new measures as well as to compare different
operationalizations of constructs and different measurement instruments.
\item By capturing underlying complex patterns and relationships,
predictive modeling can suggest improvements to existing explanatory models.
\item Scientific development requires empirically rigorous and relevant
research. Predictive modeling enables assessing the distance between
theory and practice, thereby serving as a ``reality check'' to the
relevance of theories.\footnote{Predictive models are advantageous in
terms of negative empiricism: a model either predicts accurately or it
does not, and this can be observed. In contrast, explanatory models can
never be confirmed and are harder to contradict.} While explanatory
power provides information about the strength of an underlying causal
relationship, it does not imply its predictive power.
%Only direct assessment of predictive power can shed light on the
%practical performance of explanatory models.
%
\item Predictive power assessment offers a straightforward way to
compare competing theories by examining the predictive power of their
respective explanatory models.
\item Predictive modeling plays an important role in quantifying the
level of predictability of measurable phenomena by creating benchmarks
of predictive accuracy (Ehrenberg and Bound, \citeyear{Ehrenberg1993}). Knowledge of
un-predictability is a fundamental component of scientific knowledge
(see, e.g., Taleb, \citeyear{Taleb2007}). Because predictive models tend to have
higher predictive accuracy than explanatory statistical models, they
can give an indication of the potential level of predictability. A very
low predictability level can lead to the development of new measures,
new collected data, and new empirical approaches.
%A predictive accuracy benchmark is also useful for evaluating the
%difference in predictive power of %existing explanatory models.
An explanatory model that is close to the predictive benchmark may
suggest that our understanding of that phenomenon can only be increased
marginally. On the other hand, an explanatory model that is very far
from the predictive benchmark would imply that there are substantial
practical and theoretical gains to be had from further scientific development.
\end{enumerate}

For a related, more detailed discussion of the value of prediction to
scientific theory development see the work of Shmueli and Koppius (\citeyear{ShmueliKoppius2010}).

%s1.5 ###
\subsection{Explaining and Predicting Are Different}
\label{subsec-thesame}

In the philosophy of science, it has long been debated whether
explaining and predicting are one or distinct. The conflation of
explanation and prediction has its roots in philosophy of science
literature, particularly the influential hypothetico-deductive\break model
(Hempel and Oppenheim, \citeyear{Hempel1948}), which explicitly equated prediction and explanation.
However, as later became clear, the type of uncertainty associated with
explanation is of a different nature than that associated with
prediction (Helmer and Rescher, \citeyear{Helmer1959}). This difference highlighted the need for
developing models geared specifically toward dealing with predicting
future events and trends such as the Delphi method (Dalkey and Helmer, \citeyear{Dalkey1963}).
The distinction between the two concepts has been further elaborated
(Forster and Sober, \citeyear{Forster1994}; Forster, \citeyear{Forster2002};
Sober, \citeyear{Sober2002};
Hitchcock and Sober, \citeyear{Hitchcock2004}; Dowe, Gardner and Oppy, \citeyear{Dowe2007}). In
his book \textit{Theory Building}, Dubin (\citeyear{Dubin1969}, page~9) wrote:

\begin{quote}

 Theories of social and human behavior address themselves
to two distinct goals of science: (1) prediction and (2) understanding.
It will be argued that these are separate goals [\ldots] I will not,
however, conclude that they are either inconsistent or incompatible.

\end{quote}

\noindent
Herbert Simon distinguished between ``basic science'' and ``applied
science'' (Simon, \citeyear{Simon2001}), a distinction similar to explaining versus
predicting. According to Simon, basic science is aimed at knowing (``to
describe the world'') and understanding (``to provide explanations of
these phenomena''). In contrast, in applied science, ``Laws connecting
sets of variables allow inferences or predictions to be made from known
values of some of the variables to unknown values of other variables.''

Why should there be a difference between explaining and predicting?
The answer lies in the fact that measurable data are not accurate
representations of their underlying constructs. The operationalization
of theories and constructs into statistical models and measurable data
creates a disparity between the ability to explain phenomena at the
conceptual level and the ability to generate predictions at the
measurable level.
%In the language of philosophy of science, prediction is concerned with
%regularity expressed by empirical observations; %thus empirical
%adequacy is the primary criterion of judging the efficacy of a
%predictive model. In contrast, %explanation means studying
%relationships between objects that are not directly observable; such
%relationships must be %established by theorization (\citeauthor{Yu2007}, \citeyear{Yu2007}).

%The distinction is somewhat related to the term \textit{Platonicity}
%used by \citeauthor{Taleb2007} (\citeyear{Taleb2007}) in his book \textit{The Black %Swan}
%(``Platonicity\ldots  is our tendency to mistake the map for the
%territory''.) When it comes to prediction, we are %operating on the
%level of the map, not the terrain. The two are substantially different.

To convey this disparity more formally, consider a theory postulating
that construct $\cal{X}$ causes construct $\cal{Y}$, via the function
${\cal F}$, such that ${\cal Y} = {\cal F}({\cal X})$. ${\cal F}$ is
often represented by a path model, a set of qualitative statements, a
plot (e.g., a supply and demand plot), or mathematical formulas.
Measurable variables $\mathbf{X}$ and $Y$ are operationalizations of
$\cal{X}$ and $\cal{Y}$, respectively. The operationalization of $\cal
{F}$ into a statistical model $f$, such as $E(Y) = f(\mathbf{X})$, is
done by considering ${\cal F}$ in light of the study design (e.g.,
numerical or categorical $Y$; hierarchical or flat design; time series
or cross-sectional; complete or censored data) and practical
considerations such as standards in the discipline. Because ${\cal F}$
is usually not sufficiently detailed to lead to a single $f$, often a
set of $f$ models is considered. Feelders (\citeyear{Feelders2002}) described this
process in the field of economics. In the predictive context, we
consider only $\mathbf{X}$, $Y$  and $f$.
%, as well as alternative functions $g$ that associate $\mathbf{X}$
%with $Y$.

The disparity arises because the goal in explanatory modeling is to
match $f$ and ${\cal F}$ as closely as possible for the statistical
inference to apply to the theoretical hypotheses. The data $\mathbf{X},
Y$ are tools for estimating $f$, which in turn is used for testing the
causal hypotheses. In contrast, in predictive modeling the entities of
interest are $\mathbf{X}$ and $Y$, and the function $f$ is used as a
tool for generating good predictions of new $Y$ values. In fact, we
will see that even if the underlying causal relationship is indeed
${\cal Y} = {\cal F}({\cal X})$, a function other than $\hat{f}(\mathbf
{X})$ and data other than $\mathbf{X}$ might be preferable for prediction.

The disparity manifests itself in different ways. Four major aspects are:
\begin{description}
\item[Causation--Association:] In explanatory modeling $f$ represents an
underlying causal function, and $X$ is assumed to cause $Y$. In
predictive modeling $f$ captures the association between $ \mathcal{
X}$ and ${\cal Y}$.
\item[Theory--Data:] In explanatory modeling, $f$ is carefully
constructed based on ${\cal F}$ in a fashion that supports interpreting
the estimated relationship between $X$ and $Y$ and testing the causal
hypotheses. In predictive modeling, $f$ is often constructed from the
data. Direct interpretability in terms of the relationship between $X$
and $Y$ is not required, although sometimes transparency of $f$ is desirable.
\item[Retrospective--Prospective:] Predictive modeling is
forward-looking, in that $f$ is constructed for predicting new
observations. In contrast, explanatory modeling is retrospective, in
that $f$ is used to test an already existing set of hypotheses. %In
%contrast, Although $X$ must occur chronologically before $Y$ in both
%explanatory % and predictive models, in predictive modeling an
%additional requirement is that $\mathbf{X}$ be fully %available at the
%time of prediction, if it is to be incorporated into $f$. This
%requirement is irrelevant in %explanatory modeling.
%
\item[Bias--Variance:] The expected prediction error for a new
observation with value $x$, using a quadratic loss function,\footnote
{For a binary $Y$, various 0--1 loss functions have been suggested in
place of the quadratic loss function (Domingos, \citeyear{Domingos2000}).} is given by
Hastie, Tibshirani and Friedman (\citeyear{HTF2009}, page 223)
%
%e1 ###
\begin{eqnarray}
\mathrm{EPE} %= E (Y-\hat{Y}|x)^2
&=& E \{Y-\hat{f}(x) \}^2 \nonumber\\
&=& E \{Y-f(x) \}^2 + \{E(\hat{f}(x))-f(x) \}^2 \nonumber
\\[-8pt]
\\[-8pt]
&&{}+E \{\hat{f}(x)-E(\hat
{f}(x)) \}^2 \nonumber\\
&=& \operatorname{Var}(Y) + \mathrm{Bias}^2 + \operatorname{Var}(\hat{f}(x)).\nonumber
\end{eqnarray}
Bias is the result of misspecifying the statistical model $f$. %, most
%likely due to approximating a complex reality %with a simple model.
Estimation variance (the third term) is the result of using a sample
to estimate $f$. The first term is the error that results even if the
model is correctly specified and accurately estimated. %The
%bias-variance tradeoff occurs %due to model complexity, with more
%complex models reducing bias but increasing variance (due to
%estimation error).
The above decomposition reveals a source of the difference between
explanatory and predictive modeling: In explanatory modeling the focus
is on minimizing bias to obtain the most accurate representation of the
underlying theory. In contrast, predictive modeling seeks to minimize
the combination of bias and estimation variance, occasionally
sacrificing theoretical accuracy for improved empirical precision. %
%(Hastie, Tibshirani and Friedman, \citeyear{HTF2009}, pp. 223--226).
This point is illustrated in the \hyperref[sec-appendix]{Appendix}, showing that the ``wrong''
model can sometimes predict better than the correct one.
\end{description}
%
%Another is the requirement in predictive modeling that $\mathbf{X}$ be
%fully available at the time of prediction, which %does not apply to
%explanatory modeling. We discuss these aspects in more detail in
%Section \ref{sec-process}

%, which explains why explanatory and predictive modeling warrant
%different variables (Section \ref{sec-variable}), %different methods
%(Section \ref{sec-methods}), and different model evaluation techniques
%(Section \ref{sec-eval}).

The four aspects impact every step of the modeling process, such that
the resulting $f$ is markedly different in the explanatory and
predictive contexts,
%and leading to the ``best explanatory model'' being very different
%from the ``best predictive model''
as will be shown in Section \ref{sec-process}.

%s1.6 ###
\subsection{A Void in the Statistics Literature}
\label{subsec-void}
The philosophical explaining/predicting debate has not been directly
translated into statistical language in terms of the practical aspects
of the \textit{entire} statistical modeling process. %, as shown
%schematically in Figure \ref{fig-process}.

A search of the statistics literature for discussion of explaining
versus predicting %at the practical modeling level
reveals a lively discussion in the context of \textit{model selection},
and in particular, the derivation and evaluation of model selection
criteria. In this context, Konishi and Kitagawa (\citeyear{Konishi2007}) wrote:

\begin{quote}

There may be no significant difference between the point of view of
inferring the true structure and that of making a prediction if an
infinitely large quantity of data is available or if the data are
noiseless. However, in modeling based on a finite quantity of real
data, there is a significant gap between these two points of view,
because an optimal model for prediction purposes may be different from
one obtained by estimating the `true model.'

\end{quote}

\noindent
%This discussion then leads to the argument of why AIC should be used
%when %the purpose is predictive modeling.
The literature on this topic is vast, and we do not intend to cover it
here, although we discuss the major points in Section \ref{sec-eval}.

%To the best of my knowledge, the explain/predict distinction has not
%been discussed in the more general sense of the %entire modeling
%process, beyond model selection.

The focus on prediction in the field of machine learning and by
statisticians such as Geisser, Aitchison and Dunsmore, Breiman  and
Friedman, has highlighted aspects of predictive modeling that are
relevant to the explanatory/prediction distinction, although they do
not directly contrast explanatory and predictive modeling.\footnote
{Geisser distinguished between ``[statistical] parameters'' and
``observables'' in terms of the objects of interest. His distinction is
closely related, but somewhat different from our distinction between
theoretical constructs and measurements.} The prediction literature
raises the importance of evaluating predictive power using holdout
data, and the usefulness of algorithmic methods (Breiman, \citeyear{Breiman2001}).
The predictive focus has also led to the development of inference tools
that generate predictive distributions. Geisser (\citeyear{Geisser1993}) introduced
``predictive inference'' and developed it mainly in a Bayesian context.
``Predictive likelihood'' (see Bjornstad, \citeyear{Bjornstad1990}) is a
likelihood-based approach to predictive inference, and Dawid's
prequential theory (Dawid, \citeyear{Dawid1984}) investigates inference concepts in
terms of predictability.
Finally, the bias--variance aspect has been pivotal in data mining for
understanding the predictive performance of different algorithms and
for designing new ones.

Another area in statistics and econometrics that focuses on prediction
is time series. Methods have been developed specifically for testing
the predictability of a series [e.g., random walk tests or the concept
of Granger causality (Granger, \citeyear{Granger1969})], and evaluating
predictability by examining performance on holdout data. The time
series literature in statistics is dominated by extrapolation models
such as ARIMA-type models and exponential smoothing methods, which are
suitable for prediction and description, but not for causal
explanation. Causal models for time series are common in econometrics
(e.g., Song and Witt, \citeyear{Song2000}), where an underlying causal theory links
constructs, which lead to operationalized variables, as in the
cross-sectional case. Yet, to the best of my knowledge, there is no
discussion in the statistics time series literature regarding the
distinction between predictive and explanatory modeling, aside from the
debate in economics regarding the scientific value of prediction.
%The explain/predict distinction in the time series context appears
%more straightforward, because it is tied to %chronology: forecasting
%future observations is clearly different from retrospective analysis.

%In terms of \textit{choosing potential method(s)}, a related
%discussion is Leo Breiman's paper \textit{Statistical Modeling: The
%Two Cultures} (Breiman, \citeyear{Breiman2001}, with discussion), which compares
%the statistical culture of ``models'' with the data mining culture of
%``algorithms''. The latter distinction is related to the explanatory
%vs. predictive modeling issue in terms of the choice of modeling
%tools. Whereas explanatory models usually require ``data models'',
%predictive modeling considers also ``algorithmic models'' (see Section

%Finally, in terms of comparing the development of a useful explanatory
%model to that of a powerful predictive model, although uncommon, is
%occasionally published. \cite{Athansopoulos2008} compare the
%predictive accuracy of causal explanatory models (``econometric
%models'') and predictive extrapolation models (``pure time-series
%models'') for an extensive set of tourism demand series. They conclude
%that, ``pure time-series approaches forecast tourism demand more
%accurately than the methods that use explanatory variables. This has
%immediate practical consequences, because models with explanatory
%variables are commonly used in the tourism industry and the tourism
%literature.''

To conclude, the explanatory/predictive modeling distinction has been
discussed directly in the model selection context, but not in the
larger context. Areas that focus on developing predictive modeling such
as machine learning and statistical time series, and ``predictivists''
such as Geisser, have considered prediction as a separate issue, and
have not discussed its principal and practical distinction from causal
explanation in terms of developing and testing theory.
%although some parts of the modeling process have been subject to an
%explanation/prediction lens, an overall discussion %of explanatory vs.
%predictive modeling is well needed in our discipline.
The goal of this article is therefore to examine the explanatory versus
predictive debate from a statistical perspective, considering how
modeling is used by nonstatistician scientists for theory development.

%f2 ###
\begin{figure*}

\includegraphics{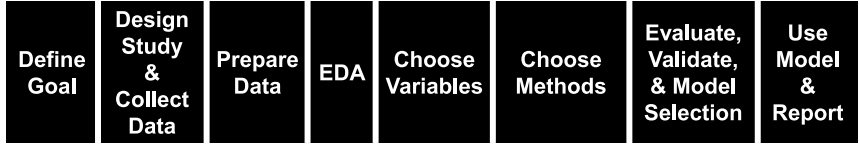}

\caption{Steps in the statistical modeling process.}
\label{fig-process}
\end{figure*}

The remainder of the article is organized as follows. In Section \ref
{sec-process}, I consider each step in the modeling process in terms of
the four aspects of the predictive/explanatory modeling distinction:
\textit{causation--\break association, theory--data,
retrospective--prospective}\break
and \textit{bias--variance}. Section \ref{sec-examples} illustrates some
of these differences via two examples. A discussion of the implications
of the predict/explain conflation, conclusions, and recommendations are
given in Section \ref{sec-conclude}.

%

%s2 ###
\section{Two Modeling Paths}
\label{sec-process}

In the following I examine the process of statistical modeling through
the explain/predict lens, from goal definition to model use and
reporting. For clarity, I~broke down the process into a generic set of
steps, as depicted in Figure \ref{fig-process}. In each step I point
out differences in the choice of methods, criteria, data, and
information to consider when the goal is predictive versus explanatory.
I also briefly describe the related statistics literature. The
conceptual and practical differences invariably lead to a difference
between a final explanatory model and a predictive one, even though
they may use the same initial data. Thus, a priori determination of the
main study goal as either explanatory or predictive\footnote{The main
study goal can also be descriptive.} is essential to conducting
adequate modeling. The discussion in this section assumes that the main
research goal has been determined as either explanatory or predictive.

%s2.1 ###
\subsection{Study Design and Data Collection}
\label{sec-design}
Even at the early stages of study design and data collection, issues of
what and how much data to collect, according to what design, and which
collection instrument to use are considered differently for prediction
versus explanation.
Consider sample size. In explanatory modeling, where the goal is to
estimate the theory-based $f$ with adequate precision and to use it for
inference, statistical power is the main consideration. Reducing bias
also requires sufficient data for model specification testing. Beyond a
certain amount of data, however, extra precision is negligible for
purposes of inference. In contrast, in predictive modeling, $f$ itself
is often determined from the data, thereby requiring a larger sample
for achieving lower bias and variance. In addition, more data are
needed for creating holdout datasets (see Section \ref
{sec-preprocess}). Finally, predicting new individual observations
accurately, in a prospective manner, requires more data than
retrospective inference regarding population-level parameters, due to
the extra uncertainty.
%A statistically significant mean effect can be practically too small
%for generating accurate individual predictions.

A second design issue is sampling scheme. For instance, in the context
of hierarchical data (e.g., sampling students within schools) Afshartous and de Leeuw (\citeyear{Afshartous2005}) noted, ``Although there exists an extensive literature
on estimation issues in multilevel models, the same cannot be said with
respect to prediction.'' Examining issues of sample size, sample
allocation, and multilevel modeling for the purpose of ``predicting a
future observable $y_{*j}$ in the $J$th group of a hierarchial
dataset,'' they found that allocation for estimation versus prediction
should be different: ``an increase in group size $n$ is often more
beneficial with respect to prediction than an increase in the number of
groups $J$\ldots  [whereas] estimation is more improved by increasing the
number of groups $J$ instead of the group size $n$.'' This relates
directly to the bias--variance aspect. A~related issue is the choice of
$f$ in relation to sampling scheme. Afshartous and de Leeuw (\citeyear{Afshartous2005}) found that
for their hierarchical data, a hierarchical $f$, which is more
appropriate theoretically, had poorer predictive performance than a
nonhierarchical $f$.
%compare three multi-level modeling methods in terms of their
%predictive accuracy.
%They find that the ``prior prediction method'', which is geared towards
%estimation and uses a hierarchical model, had %the lowest predictive
%performance, while ordinary OLS regression (where the data hierarchy
%is ignored and a flat model %is fitted) predicted nearly as accurately
%as the top method (a multilevel model with shrinkage estimators) as
%$n$ %increased.

A third design consideration is the choice between experimental and
observational settings. Whereas for causal explanation experimental
data are greatly preferred, subject to availability and resource
constraints, in prediction sometimes observational data are preferable
to ``overly clean'' experimental data, if they better represent the
realistic context of prediction in terms of the uncontrolled factors,
the noise, the measured response, etc. This difference arises from the
theory--data and prospective--retrospective aspects. Similarly, when
choosing between primary data (data collected for the purpose of the
study) and secondary data (data collected for other purposes), the
classic criteria of data recency, relevance, and accuracy (Patzer, \citeyear{Patzer1995}) are considered from a different angle. For example, a
predictive model requires the secondary data to include the exact
$\mathbf{X}, Y$ variables to be used at the time of prediction, whereas
for causal explanation different operationalizations of the constructs
${\cal X}, {\cal Y}$ may be acceptable.

In terms of the data collection instrument, whereas in explanatory
modeling the goal is to obtain a reliable and valid instrument such
that the data obtained represent the underlying construct adequately
(e.g., item response theory in psychometrics), %and on assessing
%reliability and validity of survey questions),
for predictive purposes it is more important to focus on the
measurement quality and its meaning in terms of the variable to be predicted.

Finally, consider the field of design of experiments: two major
experimental designs are factorial designs and response surface
methodology (RSM) designs. The former is focused on causal explanation
in terms of finding the factors that affect the response. The latter is
aimed at prediction---finding the combination of predictors that
optimizes $Y$. Factorial designs employ a linear $f$ for
interpretability, whereas RSM designs use optimization techniques and
estimate a nonlinear $f$ from the data, which is less interpretable but
more predictively accurate.\footnote{I thank Douglas Montgomery for
this insight.}

%s2.2 ###
\subsection{Data Preparation}
\label{sec-preprocess}

% Exploratory data analysis and data preprocessing are used
%iteratively. However, here we consider some of their aspects
%separately. We start by considering
We consider two common data preparation operations: handling missing
values and data partitioning.

%s2.2.1 ###
\subsubsection{Handling missing values}
Most real datasets consist of missing values, thereby requiring one to
identify the missing values, to determine the extent and type of
missingness, and to choose a course of action accordingly. Although a
rich literature exists on data imputation, it is monopolized by an
explanatory context.
%We begin by examining practical issues and then continue to
%methodological issues.
In predictive modeling, the solution strongly depends on whether the
missing values are in the training data and/or the data to be
predicted. For example, Sarle (\citeyear{Sarle1998}) noted:

\begin{quote}

If you have only a small proportion of cases with missing data, you can
simply throw out those cases for purposes of estimation; if you want to
make predictions for cases with missing inputs, you don't have the
option of throwing those cases out.

\end{quote}

%A second issue that Sarle raises is computational complexity, which is
%an important consideration in realtime %predictive modeling (see also
%Section \ref{sec-methods}): ``Multiple imputation, ML, and Bayesian
%methods are suitable %for prediction, but they may be too slow when
%predictions must be computed rapidly in real time.''

Sarle further listed imputation methods that are useful for explanatory
purposes but not for predictive purposes and vice versa. %explains why
%methods such as hot-deck imputation and complete case estimation are
%inapplicable in a predictive context. Moreover, he mentions that some
One example is using regression models with dummy variables that
indicate missingness, which is considered unsatisfactory in explanatory
modeling, but
%(OLS with single imputation of conditional or unconditional mean, and
%dummy variables that indicate missingness)
can produce excellent predictions.
%The latter observation about the
The usefulness of creating missingness dummy variables was also shown
by %is closely related to the work of
Ding and Simonoff (\citeyear{Ding2006}). In particular, whereas the classic explanatory
approach is based on the Missing-At-Random,
Missing-Completely-At-Random  or Not-Missing-At-Random classification
(Little and Rubin, \citeyear{Little2002}),
%, and then to choose an imputation method accordingly. F
Ding and Simonoff (\citeyear{Ding2006}) showed that for predictive purposes the important
distinction is whether the missingness depends on $Y$ or not. They
concluded:

\begin{quote}

In the context of classification trees, the relationship between the
missingness and the dependent variable, rather than the standard
missingness classification approach of Little and Rubin (2002)\ldots  is
the most helpful criterion to distinguish different missing data methods.

\end{quote}

\noindent
Moreover, missingness can be a blessing in a predictive context, if it
is sufficiently informative of $Y$ (e.g., missingness in financial
statements when the goal is to predict fraudulent reporting).

Finally, a completely different approach for handling missing data for
prediction, mentioned by Sarle (\citeyear{Sarle1998}) and further developed by
Saar-Tsechansky and Provost (\citeyear{Saar-Tsechansky2007}), considers the case %of missingness at
%prediction time,
where to-be-predicted observations are missing some predictor
information, such that the missing information can vary across
different observations. The proposed solution is %proposed the solution
%of ``reduced models'', where
to estimate multiple ``reduced'' models, %are estimated from the
%training set,
each excluding some predictors. When predicting an observation with
missingness on a certain set of predictors, the model that excludes
those predictors is used. This approach means that different reduced
models are created for different observations. Although useful for
prediction, it is clearly inappropriate for causal explanation.

%Missing data encyclopedia entry (explains Rubin's relation between
%missingness type and inference)

%s2.2.2 ###
\subsubsection{Data partitioning}
A popular solution for\break avoiding overoptimistic predictive accuracy is
to\break evaluate performance not on the training set, that is, the data used
to build the model, but rather on a holdout sample which the model
``did not see.'' The creation of a holdout sample can be achieved in
various ways, the most commonly used being a random partition of the
sample into
training and holdout sets. A popular alternative, especially with
scarce data, is cross-validation. Other alternatives are resampling
methods, such as bootstrap, which can be computationally intensive but
avoid ``bad partitions'' and enable predictive modeling with small datasets.
%Another approach to partitioning was suggested by Afshartous and de Leeuw (\citeyear{Snee1975}) who
%proposed an algorithm for creating partitions %that are as similar as
%possible.

Data partitioning is aimed at minimizing the combined bias and variance
by sacrificing some bias in return for a reduction in sampling
variance. A smaller sample is associated with higher bias when $f$ is
estimated from the data, which is common in predictive modeling but not
in explanatory modeling. Hence, data partitioning is useful for
predictive modeling but less so for explanatory modeling.
With today's abundance of large datasets, where the bias sacrifice is
practically small, data partitioning has become a standard
preprocessing step in predictive modeling.
%In explanatory modeling, where bias is the main consideration, %the
%idea of removing a holdout set from data at hand
%is typically unacceptable.
%However, with a large dataset the reduction in sample size for the
%training set will not be substantial and in today's %data environment,
% Data partitioning is also important in predictive modeling because it
%allows evaluating predictive performance when %assumptions about error
%distributions are relaxed.

In explanatory modeling, data partitioning is less common because of
the reduction in statistical power. When used, it is usually done for
the retrospective purpose of assessing the robustness of $\hat{f}$. A
rarer yet important use of data partitioning in explanatory modeling is
for strengthening model validity, by demonstrating some predictive
power. Although one would not expect an explanatory model to be optimal
in terms of predictive power, it should show some degree of accuracy
(see discussion in Section \ref{subsec-2dim}).

%Finally, although data partitioning is most suited to large datasets,
%alternatives exist specifically for small datasets, such as
%cross-validation or resampling methods such as bootstrap. Predictive
%modeling can thus be used effectively in small datasets.

%s2.3 ###
\subsection{Exploratory Data Analysis}
\label{sec-explore}

Exploratory data analysis (EDA) is a key initial step in both
explanatory and predictive modeling. It consists of summarizing the
data numerically and graphically, reducing their dimension, and
``preparing'' for the more formal modeling step. Although the same set
of tools can be used in both cases, they are used in a different
fashion. In explanatory modeling, %often termed ``confirmatory
%analysis'', the
exploration is channeled toward the theoretically specified causal
relationships, whereas %. In contrast, the goal
in predictive modeling EDA is used in a more free-form fashion,
supporting the purpose of capturing relationships that are perhaps
unknown or at least less formally formulated.% Thus, EDA is used in the
%predictive context in a more %free-form fashion.

One example is how data visualization is carried out. Fayyad, Grinstein and Wierse
(\citeyear{Fayyad2002}, page~22) contrasted ``exploratory visualization'' with
``confirmatory visualization'':

\begin{quote}

Visualizations can be used to explore data, to confirm a hypothesis, or
to manipulate a viewer\ldots  In exploratory visualization the user does
not necessarily know what he is looking for. This creates a dynamic
scenario in which interaction is critical\ldots
% The user is searching for structure or trends and is attempting to
%arrive at some hypothesis.
In a confirmatory visualization, the user has a hypothesis that needs
to be tested.
This scenario is more stable and predictable. System parameters are
often predetermined.

\end{quote}

\noindent
Hence, \textit{interactivity}, which supports exploration\break across a wide
and sometimes unknown terrain, is very useful for learning about
measurement quality and associations that are at the core of predictive
modeling, but much less so in explanatory modeling, where the data are
visualized through the theoretical lens.

A second example is numerical summaries. In a predictive context, one
might explore a wide range of numerical summaries for all variables of
interest, whereas in an explanatory model, the numerical summaries
would focus on the theoretical relationships. For example, in order to
assess the role of a certain variable as a mediator, its correlation
with the response variable and with other covariates is examined by
generating specific correlation tables.

A third example is the use of EDA for assessing assumptions of
potential models (e.g., normality or multicollinearity) and exploring
possible variable transformations. Here, too, an explanatory context
would be more restrictive in terms of the space explored.

Finally, dimension reduction is viewed and used differently. In
predictive modeling, a reduction in the number of predictors can help
reduce sampling variance. Hence, methods such as principal components
analysis (PCA) or other data compression methods that are even less
interpretable (e.g., singular value decomposition) are often carried
out initially. They may later lead to the use of compressed variables
(such as the first few components) as predictors, even if those are not
easily interpretable. PCA is also used in explanatory modeling, but for
a different purpose. For questionnaire data, PCA and exploratory factor
analysis are used to determine the validity of the survey instrument.
The resulting factors are expected to correspond to the underlying constructs.
%, thereby supporting the validity of the instrument.
In fact, the rotation step in factor analysis is specifically aimed at
making the factors more interpretable. Similarly, correlations are used
for assessing the reliability of the survey instrument.

%s2.4 ###
\subsection{Choice of Variables}
\label{sec-variable}

The criteria for choosing variables differ markedly in explanatory
versus predictive contexts.

In explanatory modeling, where variables are seen as operationalized
constructs, variable choice is based on the role of the construct in
the theoretical causal structure and on the operationalization itself.
A broad terminology related to different variable roles exists in
\mbox{various} fields:
% One example is the special attention in the social sciences paid to
%defining the type of variable:
in the social sciences---\textit{antecedent, consequent, mediator}
and \textit{moderator}\footnote{``A moderator variable is one that
influences the strength of a relationship between two other variables,
and a mediator variable is one that explains the relationship between
the two other variables'' (from
\url{http://psych.wisc.edu/henriques/mediator.html}).} variables; in
pharmacology and medical sciences---\textit{treatment} and \textit
{control} variables; and in epidemiology---\textit{exposure} and \textit
{confounding} variables. Carte and Craig (\citeyear{Carte2003}) mentioned that explaining
moderating effects has become an important scientific endeavor in the
field of Management Information Systems.
%(MIS): ``The increasing interest in moderated relationships
%reinforces a notion that MIS researchers are increasingly %addressing:
%context matters in MIS research.''
Another important term common in economics is \textit{endogeneity} or
``reverse causation,'' which results in biased parameter estimates.
Endogeneity can occur due to different reasons. One reason is
incorrectly omitting an input variable, say $Z$, from $f$ when the
causal construct ${\cal Z}$ is assumed to cause ${\cal X}$ and ${\cal
Y}$. In a regression model of $Y$ on $X$, the omission of $Z$ results
in $X$ being correlated with the error term. Winkelmann (\citeyear{Winkelmann2008})
gave the example of a hypothesis that health insurance (${\cal X}$)
affects the demand for health services ${\cal Y}$. The operationalized
variables are ``health insurance status'' ($X$) and ``number of doctor
consultations'' ($Y$). Omitting an input measurement $Z$ for ``true
health status'' (${\cal Z}$) from the regression model $f$ causes
endogeneity because $X$ can be determined by $Y$ (i.e., reverse
causation), which manifests as $X$ being correlated with the error term
in $f$. Endogeneity can arise due to other reasons such as measurement
error in $X$.
%An endogenous variable $X$ is assumed to be caused by other variables
%in the theoretical model. In a regression model, for instance, an
%endogenous covariate is one that correlates with the error term,
%indicating the possibility that the response causes the covariate.
%Endogeneity, in this case, is attributed to omitted covariates or to
%measurement error, leading to biased parameters estimates.
Because of the focus in explanatory modeling on causality and on bias,
there is a vast literature on detecting endogeneity and on solutions
such as constructing instrumental variables and using models such as
two-stage-least-squares (2SLS). Another related term is \textit
{simultaneous causality}, which gives rise to special models such as
Seemingly Unrelated Regression (SUR) (Zellner, \citeyear{Zellner1962}). In terms of
chronology, a causal explanatory model can include only ``control''
variables that take place before the causal variable (Gelman et al., \citeyear{Gelman2003}).
And finally, for reasons of model identifiability (i.e., given the
statistical model, each causal effect can be identified), one is
required to include main effects in a model that contains an
interaction term between those effects. We note this practice because
it is not necessary or useful in the predictive context, due to the
acceptability of uninterpretable models and the potential reduction in
sampling variance when dropping predictors (see, e.g., the \hyperref[sec-appendix]{Appendix}).

In predictive modeling, the focus on association rather than causation,
the lack of ${\cal F}$, and the prospective context, mean that there is
no need to delve into the exact role of each variable in terms of an
underlying causal structure. Instead, criteria for choosing predictors
are quality of the association between the predictors and the response,
data quality, and availability of the predictors at the time of
prediction, known as ex-ante availability.
%To see the extent of the practical and theoretical gap between
%explanation and prediction, let us contrast variable %choice on the
%criterion of ex-ante availability.
In terms of ex-ante availability, whereas chronological precedence of
$X$ to $Y$ is necessary in causal models, in predictive models not only
must $X$ precede $Y$, but $X$ must be available at the time of
prediction. For instance, explaining wine quality retrospectively would
dictate including barrel characteristics as a causal factor. The
inclusion of barrel characteristics in a predictive model of future
wine quality would be impossible if at the time of prediction the
grapes are still on the vine. See the eBay example in Section \ref
{subsec-eBay} for another example.
%In econometrics, the terms \textit{leading, lagging,} and
%variable relative to an event of interest (before, after, or
%concurrent, respectively).
%For example, consider the explanation vs. prediction of the final
%price in a auction. Game theory suggests that one of %the factors
%affecting the final price is the number of bidders (Krishna, \citeyear{kris2002}).
%However, if our goal is to forecast %the price of an ongoing auction
%where the number of bidders is unknown until the auction closes (as is
%the case on %eBay), a model that contains the number of bidders as a
%predictor is useless. In comparison, such an explanatory model %can be
%useful for retrospectively studying the effect of the number of
%bidders on the auction price.

%s2.5 ###
\subsection{Choice of Methods}
\label{sec-methods}

Considering the four aspects of causation--associa\-tion, theory--data,
retrospective--prospective  and bias--variance leads to different choices
of plausible methods, with a much larger array of methods useful for
prediction. Explanatory modeling requires interpretable statistical
models $f$ that are easily linked to the underlying theoretical model
${\cal F}$. Hence the popularity of statistical models, and especially
regression-type methods, in many disciplines. Algorithmic methods such
as neural networks or $k$-nearest-neighbors, and uninterpretable
nonparametric models, are considered ill-suited for explanatory modeling.

In predictive modeling, where the top priority is generating accurate
predictions of new observations and $f$ is often unknown, the range of
plausible methods includes not only statistical models (interpretable
and uninterpretable) but also data mining algorithms. A~neural network
algorithm might not shed light on an underlying causal mechanism ${\cal
F}$ or even on $f$, but it can capture complicated associations,
thereby leading to accurate predictions. Although model\break transparency
might be important in some cases, it is of secondary importance:
``Using complex predictors may be unpleasant, but the soundest path is
to go for predictive accuracy first, then try to understand why'' (Breiman, \citeyear{Breiman2001}).

Breiman (\citeyear{Breiman2001}) accused the statistical community of ignoring
algorithmic modeling:

\begin{quote}

There are two cultures in the use of statistical modeling to reach
conclusions from data. One assumes that the data are generated by a
given stochastic data model. The other uses algorithmic models and
treats the data mechanism as unknown. The statistical community has
been committed to the almost exclusive use of data models.

\end{quote}

\noindent
From the explanatory/predictive view, algorithmic\break  modeling is indeed
very suitable for predictive (and descriptive) modeling, but not for
explanatory modeling.

Some methods are not suitable for prediction\break from the
retrospective--prospective aspect, especially in time series
forecasting. Models with coincident indicators, which are measured
simultaneously, are such a class. An example is the model $\mathit{Airfare}_t =
f(\mathit{OilPrice}_t)$, which might be useful for explaining the effect of oil
price on airfare based on a causal theory, but not for predicting
future airfare because the oil price at the time of prediction is
unknown. For prediction, alternative models must be considered, such as
using a lagged \textit{OilPrice} variable, or creating a separate model
for forecasting oil prices and plugging its forecast into the airfare
model. Another example is the centered moving average, which requires
the availability of data during a time window before and after a period
of interest, and is therefore not useful for prediction.

Lastly, the bias--variance aspect raises two classes of methods that are
very useful for prediction, but not for explanation. The first is
shrinkage methods such as ridge regression, principal components
regression, and partial least squares regression, which ``shrink''
predictor coefficients or even eliminate them, thereby introducing bias
into $f$, for the purpose of reducing estimation variance. The second
class of methods, which ``have been called the most influential
development in Data Mining and Machine Learning in the past decade''
(Seni and Elder, \citeyear{seni2010}, page vi), are ensemble methods such as bagging\break
(Breiman, \citeyear{Breiman1996}), random forests (Breiman, \citeyear{Breiman2001a}),
boosting\footnote
{Although boosting algorithms were developed as ensemble methods,
``[they can] be seen as an interesting regularization scheme for
estimating a model'' (Bohlmann and Hothorn, \citeyear{Bohlmann2007}).} (Schapire, \citeyear{Shapire1999}),
variations of those methods, and Bayesian alternatives (e.g., Brown, Vannucci and Fearn, \citeyear{Brown2002}). Ensembles combine multiple models to produce more
precise predictions by averaging predictions from different models, and
have proven useful in numerous applications (see the Netflix Prize
example in Section \ref{subsec-netflix}).

%s2.6 ###
\subsection{Validation, Model Evaluation  and Model Selection}
\label{sec-eval}

Choosing the final model among a set of models, validating it, and
evaluating its performance, differ markedly in explanatory and
predictive modeling. Although the process is iterative, I separate it
into three components for ease of exposition.

%s2.6.1 ###
\subsubsection{Validation}

In explanatory modeling, validation consists of two parts: \textit
{model validation} validates that $f$ adequately represents ${\cal F}$,
and \textit{model fit} validates that $\hat{f}$ fits the data $\{X, Y\}
$. In contrast, validation in predictive modeling is focused on \textit
{generalization}, which is the ability of $\hat{f}$ to predict new data
$\{X_{\mathrm{new}}, Y_{\mathrm{new}}\}$.

Methods used in explanatory modeling for model validation include model
specification tests such as the popular Hausman specification test in
econometrics (Hausman, \citeyear{Hausman1978}), and construct validation techniques
such as reliability and validity measures of survey questions and
factor analysis. Inference for individual coefficients is also used for
detecting over- or underspecification. Validating model fit involves
goodness-of-fit tests (e.g., normality tests) and model diagnostics
such as residual analysis. Although indications of lack of fit might
lead researchers to modify $f$, modifications are made carefully in
light of the relationship with ${\cal F}$ and the constructs ${\cal X, Y}$.

In predictive modeling, the biggest danger to generalization is
overfitting the training data. Hence validation consists of evaluating
the degree of\vspace*{1pt} overfitting, by comparing the performance of $\hat{f}$ on
the training and holdout sets. If performance is significantly better
on the training set, overfitting is implied.

Not only is the large context of validation markedly different in
explanatory and predictive modeling, but so are the details. For
example, checking for multicollinearity is a standard operation in
assessing model fit. This practice is relevant in explanatory modeling,
where multicollinearity
%Although most introductory statistics textbooks mention
%multicollinearity as a major danger, the importance of %detecting and
%treating multicollinearity is different in explanation vs. prediction.
%In explanatory models, %multicollinearity is considered a ``danger''
%because it
can lead to inflated standard errors, which interferes with inference.
Therefore, a vast literature exists on strategies for identifying and
reducing multicollinearity, variable selection being one strategy. In
contrast, for predictive purposes ``multicollinearity is not quite as
damning'' (Vaughan and Berry, \citeyear{Vaughan2005}). Makridakis, Wheelwright and Hyndman (\citeyear{Makridakis1997},
page~288)
distinguished between the role of multicollinearity in explaining
versus its role in predicting:

\begin{quote}

 Multicollinearity is not a problem unless either (i) the
individual regression coefficients are of interest, or (ii) attempts
are made to isolate the contribution of one explanatory variable to Y,
\textit{without} the influence of the other explanatory variables.
Multicollinearity will not affect the ability of the model to predict.

\end{quote}

Another example is the detection of influential observations. While
classic methods are aimed at detecting observations that are
influential in terms of model estimation, Johnson and Geisser (\citeyear{Johnsone1983}) proposed
a method for detecting influential observations in terms of their
effect on the predictive distribution.

%s2.6.2 ###
\subsubsection{Model evaluation}
\label{subsec-performance}

Consider two performance aspects of a model: explanatory power and
predictive power.
The top priority in terms of model performance in explanatory modeling
is assessing \textit{explanatory power}, which measures the \textit
{strength of relationship} indicated by $\hat{f}$. Researchers report
$R^2$-type values and statistical significance of overall $F$-type
statistics to indicate the level of explanatory power.

In contrast, in predictive modeling, the focus is on \textit{predictive
accuracy} or \textit{predictive power}, which refer to the performance
of $\hat{f}$ on new data. Measures of predictive power are typically
out-of-sample metrics or their in-sample approximations, which depend
on the type of required prediction. For example, predictions of a
binary $Y$ could be binary classifications ($\hat{Y}=0,1$), predicted
probabilities of a certain class [$\hat{P}(Y=1)$], or rankings of those
probabilities. The latter are common in marketing and personnel
psychology. These three different types of predictions would warrant
different performance metrics. For example, a model can perform poorly
in producing binary classifications but adequately in producing
rankings. Moreover, in the context of asymmetric costs, where costs are
heftier for some types of prediction errors than others, alternative
performance metrics are used, such as the ``average cost per predicted
observation.''

A common misconception in various scientific fields is that predictive
power can be inferred from explanatory power. However, the two are
different and should be assessed separately. While predictive power can
be assessed for both explanatory and predictive models, explanatory
power is not typically possible to assess for predictive models because
of the lack of ${\cal F}$ and an underlying causal structure. Measures
such as $R^2$ and $F$ would indicate the level of association, but not
causation.

Predictive power is assessed using metrics computed from a holdout set
or using cross-validation (Stone, \citeyear{Stone1974};
Geisser, \citeyear{Geisser1975}). Thus, a major
difference between explanatory and predictive performance metrics is
\textit{the data from which they are computed}.
In general, measures computed from the data to which the model was
fitted tend to be overoptimistic in terms of predictive accuracy:
``Testing the procedure on the data that gave it birth is almost
certain to overestimate performance'' (Mosteller and Tukey, \citeyear{MostellerTukey1977}). Thus,
the holdout set serves as a more realistic context for evaluating
predictive power.

%s2.6.3 ###
\subsubsection{Model selection}

Once a set of models $f_1, f_2,\break  \ldots $ has been estimated and
validated, model selection pertains to choosing among them. Two main
differentiating aspects are the data--theory and bias--variance
considerations. In explanatory modeling, the models are compared in
terms of explanatory power, and hence the popularity of nested models,
which are easily compared. Stepwise-type methods, which use overall $F$
statistics to include and/or exclude variables, might appear suitable
for achieving high explanatory power. However, optimizing explanatory
power in this fashion conceptually contradicts the validation step,
where variable inclusion/exclusion and the structure of the statistical
model are carefully designed to represent the theoretical model. Hence,
proper explanatory model selection is performed in a constrained
manner. In the words of Jaccard (\citeyear{Jaccard2001}):

\begin{quote}

 Trimming potentially theoretically meaningful variables
is not advisable unless one is quite certain that the coefficient for
the variable is near zero, that the variable is inconsequential, and
that trimming will not introduce misspecification error.

\end{quote}

A researcher might choose to retain a causal covariate which has a
strong theoretical justification \textit{even if is statistically
insignificant}. For example, in medical research, a covariate that
denotes whether a person smokes or not is often present in models for
health conditions, whether it is statistically significant or
not.\footnote{I thank Ayala Cohen for this example.} In contrast to
explanatory power, statistical significance plays a minor or no role in
assessing predictive performance. In fact, it is sometimes the case
that removing inputs with small coefficients, \textit{even if they are
statistically significant}, results in improved prediction accuracy
(Greenberg and Parks, \citeyear{Greenberg1997}; Wu, Harris and McAuley,
\citeyear{Wu2007}, and see the \hyperref[sec-appendix]{Appendix}). Stepwise-type
algorithms are very useful in predictive modeling as long as the
selection criteria rely on predictive power rather than explanatory power.

As mentioned in Section \ref{subsec-void}, the statistics literature on
model selection includes a rich discussion on the difference between
finding the ``true'' model and finding the best predictive model, and on
criteria for explanatory model selection versus predictive model
selection. A popular predictive metric is the in-sample Akaike
Information Criterion (AIC). Akaike derived the AIC from a predictive
viewpoint, where the model is not intended to accurately infer the
``true distribution,'' but rather to predict future data as accurately
as possible (see, e.g., Berk, \citeyear{Berk2008};
Konishi and Kitagawa, \citeyear{Konishi2007}). Some researchers
distinguish between AIC and the Bayesian information criterion (BIC) on
this ground. Sober (\citeyear{Sober2002}) concluded that AIC measures predictive
accuracy while BIC measures goodness of fit:

\begin{quote}

 In a sense, the AIC and the BIC provide estimates of
different things; yet, they almost always are thought to be in
competition. If the question of which estimator is better is to make
sense, we must decide whether the average likelihood of a family [$=$BIC]
or its predictive accuracy [$=$AIC] is what we want to estimate.

\end{quote}

Similarly, Dowe, Gardner and Oppy (\citeyear{Dowe2007}) contrasted the two Bayesian model selection
criteria Minimum Message Length (MML) and Minimum Expected
Kullback--Leibler Distance (MEKLD).\break They concluded,

\begin{quote}

 If you want to maximise predictive accuracy, you should
minimise the expected KL distance (MEKLD); if you want the best
inference, you should use MML.

\end{quote}

\noindent
Kadane and Lazar (\citeyear{Kadane2004}) examined a variety of model selection criteria from
a Bayesian decision--theoretic point of view, comparing prediction with
explanation goals.

Even when using predictive metrics, the fashion in which they are used
within a model selection process can deteriorate their adequacy,
yielding overoptimistic predictive performance. Berk (\citeyear{Berk2008})
described the case where

\begin{quote}

 statistical learning procedures are often applied several
times to the data with one or more tuning parameters varied. The AIC
may be computed for each. But each AIC is ignorant about the
information obtained from prior fitting attempts and how many degrees
of freedom were expended in the process. Matters are even more
complicated if some of the variables are transformed or recoded\ldots  Some
unjustified optimism remains.

\end{quote}

%s2.7 ###
\subsection{Model Use and Reporting}
\label{sec-deploy}

Given all the differences that arise in the modeling process, the
resulting predictive model would obviously be very different from a
resulting explanatory model in terms of the data used ($\{X, Y\}$), the
estimated model $\hat{f}$, and explanatory power and predictive power.
The use of $\hat{f}$ would also greatly differ.

As illustrated in Section \ref{subsec-explain}, explanatory models in
the context of scientific research are used to derive ``statistical
conclusions'' using inference, which in turn are translated into
scientific conclusions regarding ${\cal F,X,Y}$ and the causal
hypotheses. With a focus on theory, causality, bias and retrospective
analysis, explanatory studies are aimed at testing or comparing
existing causal theories. Accordingly the statistical section of
explanatory scientific papers is dominated by statistical inference.
%In addition to statistical significance and effect interpretation,
%causal diagrams are often used (such as Figure %
%theoretical model and due to the heavy reliance on %classical
%statistical inference, the empirical model is evaluated in terms of
%its relation to some ``null'' model.

In predictive modeling $\hat{f}$ is used to generate predictions for
new data. We note that generating predictions from $\hat{f}$ can range
in the level of difficulty, depending on the complexity of $\hat{f}$
and on the type of prediction generated. For example, generating a
complete predictive distribution is easier using a\break Bayesian approach
than the predictive likelihood approach.

In practical applications, the predictions might be the final goal.
However, the focus here is on predictive modeling for supporting
scientific research, as was discussed in Section \ref{subsec-predict}.
Scientific predictive studies and articles therefore emphasize data,
association, bias--variance considerations, and prospective aspects of
the study. Conclusions pertain to theory-building aspects such as new
hypothesis generation, practical relevance, and predictability level.
Whereas explanatory articles focus on theoretical constructs and
unobservable parameters and their statistical section is dominated by
inference, predictive articles concentrate on the observable level,
with predictive power and its comparison across models being the core.

%Scientific reports of a predictive nature focus on the observable
%level. Data summarization and visualization is typically presented in
%more detail than in explanatory studies, and attention is given to the
%various technical details of the different modeling steps, justified
%by improvements in predictive performance. Performance is reported
%numerically and graphically in terms of the predictive accuracy of
%interest (e.g., confusion matrices, overall accuracy or cost measures,
%lift charts and ROC curves). Model performance is typically compared
%against a naive model and alternative predictive models, both for the
%training and for the holdout data. In addition, the treatment of
%over-fitting is often discussed. Depending on the scientific goal(s)
%of the study (as discussed above), the results would translate into
%scientific conclusions.

%s3 ###
\section{Two Examples}
\label{sec-examples}

Two examples are used to broadly illustrate the differences that arise
in predictive and explanatory studies. In the first I consider a
predictive goal and discuss what would be involved in ``converting'' it
to an explanatory study. In the second example I consider an
explanatory study and what would be different in a predictive context.
See the work of Shmueli and Koppius (\citeyear{ShmueliKoppius2010}) for a detailed example
``converting'' the explanatory study of Gefen, Karahanna and Straub (\citeyear{Gefen2003}) from Section
\ref{sec-intro} into a predictive one.

%s3.1 ###
\subsection{Netflix Prize}
\label{subsec-netflix}

Netflix is the largest online DVD rental service in the United States.
In an effort to improve their movie recommendation system, in 2006
Netflix announced a contest (\url{http://netflixprize.com}), making public a
huge dataset of user movie ratings. Each observation consisted of a
user ID, a movie title, and the rating that the user gave this movie.
The task was to accurately predict the ratings of movie-user pairs for
a test set such that the predictive accuracy improved upon Netflix's
recommendation engine by at least 10\%. The grand prize was set at \$
1,000,000. The 2009 winner was a composite of three teams, one of them
from the AT\&T research lab (see Bell, Koren and Volinsky, \citeyear{Bell2010}). In their 2008
report, the AT\&T team, who also won the 2007 and 2008 progress prizes,
described their modeling approach (Bell, Koren and Volinsky, \citeyear{Bell2008}).

Let me point out several operations and choices described by Bell, Koren and Volinsky (\citeyear{Bell2008}) that highlight the distinctive predictive context. Starting
with sample size, the very large sample released by Netflix was aimed
at allowing the estimation of $f$ from the data, reflecting the absence
of a strong theory. In the data preparation step, with relation to
missingness that is predictively informative, the team found that ``the
information on which movies each user chose to rate, regardless of
specific rating value'' turned out to be useful. At the data exploration
and reduction step, many teams including the winners found that the
noninterpretable Singular Value Decomposition (SVD) data reduction
method was key in producing accurate predictions: ``It seems that
models based on matrix-factorization were found to be most accurate.''
As for choice of variables, supplementing the Netflix data with
information about the movie (such as actors, director) actually
decreased accuracy: ``We should mention that not all data features were
found to be useful. For example, we tried to benefit from an extensive
set of attributes describing each of the movies in the dataset. Those
attributes certainly carry a significant signal and can explain some of
the user behavior. However, we concluded that they could not help at
all for improving the accuracy of well tuned collaborative filtering models.''
In terms of choice of methods, their solution was an ensemble of
methods that included nearest-neighbor algorithms, regression models,
and shrinkage methods. In particular, they found that ``using
increasingly complex models is only one way of improving accuracy. An
apparently easier way to achieve better accuracy is by blending
multiple simpler models.'' And indeed, more accurate predictions were
achieved by collaborations between competing teams who combined
predictions from their individual models, such as the winners' combined
team. All these choices and discoveries are very relevant for
prediction, but not for causal explanation. Although the Netflix
contest is not aimed at scientific advancement, there is clearly
scientific value in the predictive models developed. They tell us about
the level of predictability of online user ratings of movies, and the
implicated usefulness of the rating scale employed by Netflix. The
research also highlights the importance of knowing which movies a user
does not rate. And importantly, it sets the stage for explanatory research.

Let us consider a hypothetical goal of \textit{explaining} movie
preferences. After stating causal hypotheses, we would define
constructs that link user behavior and movie features $ {\mathcal{
X}}$
to user preference ${\cal Y}$, with a careful choice of ${\cal F}$.
%reflecting the causal structure.
An operationalization step would link the constructs to measurable
data, and the role of each variable in the causality structure would be
defined. Even if using the Netflix dataset, supplemental covariates
that capture movie features and user characteristics would be
absolutely necessary. In other words, the data collected and the
variables included in the model would be different from the predictive
context. As to methods and models, data compression methods such as
SVD, heuristic-based predictive algorithms which learn $f$ from the
data, and the combination of multiple models would be considered
inappropriate, as they lack interpretability with respect to ${\cal F}$
and the hypotheses. The choice of $f$ would be restricted to
statistical models that can be used for inference, and would directly
model issues such as the dependence between records for the same
customer and for the same movie. Finally, the model would be validated
and evaluated in terms of its explanatory power, and used to conclude
about the strength of the causal relationship between various user and
movie characteristics and movie preferences. Hence, the explanatory
context leads to a completely different modeling path and final result
than the predictive context.

It is interesting to note that most competing teams had a background in
computer science rather than statistics. Yet, the winning team combines
the two disciplines. Statisticians who see the uniqueness and
importance of predictive modeling alongside explanatory modeling have
the capability of contributing to scientific advancement as well as
achieving meaningful practical results (and large monetary awards).

%s3.2 ###
\subsection{Online Auction Research}
\label{subsec-eBay}

The following example highlights the differences between explanatory
and predictive research in online auctions. The predictive approach
also illustrates the utility in creating new theory in an area
dominated by explanatory modeling.

Online auctions have become a major player in providing electronic
commerce services. eBay (\href{http://www.eBay.com}{www.}\break
\href{http://www.eBay.com}{eBay.com}), the largest
consumer-to-consumer auction website, enables a global community of
buyers and sellers to easily interact and trade. Empirical research of
online auctions has grown dramatically in recent years. Studies using
publicly available bid data from websites such as eBay have found many
divergences of bidding behavior and auction outcomes compared to
ordinary offline auctions and classical auction theory. For instance,
according to classical auction theory (e.g., Krishna, \citeyear{kris2002}), the
final price of an auction is determined by a priori information about
the number of bidders, their valuation, and the auction format.
However, final price determination in online auctions is quite
different. Online auctions differ from offline auctions in various ways
such as longer duration, anonymity of bidders and sellers, and low
barriers of entry. These and other factors lead to
%dynamics in the bid arrival and price process that change throughout
%the auction, and to
new bidding behaviors that are not explained by auction theory. Another
important difference is that the total number of bidders in most online
auctions is unknown until the auction closes.

Empirical research in online auctions has concentrated in the fields of
economics, information systems and marketing. Explanatory modeling has
been employed to learn about different aspects of bidder behavior in
auctions. A survey of empirical explanatory research on auctions was
given by Bajari and Hortacsu (\citeyear{bajahort2004}). A typical explanatory study relies on
game theory to construct ${\cal F}$, which can be done in different
ways. One approach is to construct a ``structural model,'' which is a
mathematical model linking the various constructs. The major construct
is ``bidder valuation,'' which is the amount a bidder is willing to pay,
and is typically operationalized using his observed placed bids. The
structural model and operationalized constructs are then translated
into a regression-type model [see, e.g., Sections 5 and 6 in Bajari and Hortacsu (\citeyear{bajahort2003})].
To illustrate the use of a statistical model in
explanatory auction research, consider the study by Lucking-Reiley et al. (\citeyear{Reiley2007})
who used a dataset of 461 eBay coin auctions to determine the factors
affecting the final auction price. They estimated a set of linear
regression models where $Y=\log(\mathit{Price})$ and $X$ included auction
characteristics (the opening bid, the auction duration, and whether a
secret reserve price was used), seller characteristics (the number of
positive and negative ratings), and a control variable (book value of
the coin). One of their four reported models was of the form
\begin{eqnarray*}
\log(\mathit{Price}) &=& \beta_0 + \beta_1 \log(\mathit{BookValue})\\
&&{}+ \beta_2 \log(\mathit{MinBid})
+ \beta_3 \mathit{Reserve} \\
&&{}+ \beta_4 \mathit{NumDays} + \beta_5 \mathit{PosRating}\\
&&{}+ \beta_6 \mathit{NegRating} +\epsilon.
\end{eqnarray*}
The other three models, or ``model specifications,'' included a modified
set of predictors, with some interaction terms and an alternate auction
duration measurement. The authors used a censored-Normal regression for
model estimation, because some auctions did not receive any bids and
therefore the price was truncated at the minimum bid. Typical
explanatory aspects of the modeling are:
\begin{description}
\item[Choice of variables:] Several issues arise from the\break
causal-theoretical context. First is the exclusion of the number of
bidders (or bids) as a determinant due to endogeneity considerations,
where although it is likely to affect the final price, ``it is
endogenously determined by the bidders' choices.'' To verify endogeneity
the authors report fitting a separate regression of \textit{Y${}={} $Number
of bids} on all the determinants. Second, the authors discuss
operationalization challenges that might result in bias due to omitted
variables. In particular, the authors discuss the construct of
``auction attractiveness'' (${\cal X}$) and their inability to judge
measures such as photos and verbal descriptions to operationalize
attractiveness.
\item[Model validation:] The four model specifications are used for
testing the robustness of the hypothesized effect of the construct
``auction length'' across different operationalized variables such as
the continuous number of days and a categorical alternative.
\item[Model evaluation:] For each model, its in-sample $R^2$ is used
for determining explanatory power.
\item[Model selection:] The authors report the four fitted regression
models, including both significant and insignificant coefficients.
Retaining the insignificant covariates in the model is for matching $f$
with~${\cal F}$.
\item[Model use and reporting:] The main focus is on inference for the
$\beta$'s, and the final conclusions are given in causal terms. (``A
seller's feedback ratings\ldots  have a measurable effect on her auction
prices\ldots  when a seller chooses to have her auction last for a
longer period of days [sic], this significantly increases the auction
price on average.'')
\end{description}

Although online auction research is dominated by explanatory studies,
there have been a few predictive studies developing forecasting models
for an auction's final price
(e.g., Jank, Shmueli and Wang, \citeyear{Jank2008}; Jap and Naik, \citeyear{Jap2008};
Ghani and Simmons, \citeyear{Ghani2004};
Wang, Jank and Shmueli, \citeyear{WangJankShmueli2008};
Zhang, Jank and Shmueli, \citeyear{Zhang2009}).
For a brief survey of online auction forecasting research see the work
of Jank and Shmueli (\citeyear{JankShmueli2010}, Chapter~5). From my involvement in several
of these predictive studies, let me highlight the purely predictive
aspects that appear in this literature:
\begin{description}
\item[Choice of variables:] If prediction takes place before or at the
start of the auction, then obviously the total number of bids or
bidders cannot be included as a predictor. While this variable was also
omitted in the explanatory study, the omission was due to a different
reason, that is, endogeneity. However, if prediction takes place at
time $t$ during an ongoing auction, then the number of bidders/bids
present at time $t$ is available and useful for predicting the final
price. Even more useful is the time series of the number of bidders
from the start of the auction until time $t$ as well as the price curve
until time $t$ (Bapna, Jank and Shmueli, \citeyear{BapnaJankShmueli2008}).
\item[Choice of methods:] Predictive studies in online auctions tend to
learn $f$ from the data, using flexible models and algorithmic methods
(e.g., CART, $k$-nearest neighbors, neural networks, functional methods
and related nonparametric smoothing-based methods, Kalman filters  and
boosting (see, e.g., Chapter~5 in Jank and Shmueli, \citeyear{JankShmueli2010}).
Many of
these are not interpretable, yet have proven to provide high predictive
accuracy.
\item[Model evaluation:] Auction forecasting studies evaluate
predictive power on holdout data. They report performance in terms of
out-of-sample metrics such as \textit{MAPE} and \textit{RMSE}, and are
compared against other predictive models and benchmarks.
\end{description}

%The predictive studies rely on holdout predictions for performance
%evaluation; predictors are selected and transformed with predictive
%accuracy and ex-ante availability in mind (e.g., some methods include
%dynamically updated information that takes place during the live
%auction), types of models include very flexible models and algorithmic
%methods (e.g., CART, $k$-nearest neighbors, neural networks,
%functional methods and related non-parametric smoothing-based methods,
%and Kalman filters). Model selection and evaluation are based on
%predictive performance and comparisons with alternative prediction
%methods.

Predictive models for auction price cannot provide direct causal
explanations. However, by producing high-accuracy price predictions
they shed light on new potential variables that are related to price
and on the types of relationships that can be further investigated in
terms of causality. For instance, a construct that is not directly
measurable but that some predictive models are apparently capturing is
competition between bidders. %Another leap from % prediction to
%explanation can be obtained by studying predictive accuracy in
%different contexts. For example, the % method based on functional data
%modeling by Wang, Jank and Shmueli (\citeyear{WangJankShmueli2008}) has been shown to provide high
%predictive %accuracy for various types of items in various online
%auction markets \citep{Dass2009, Jank2006}.

%
%CHEMICAL ENGINEERING - WU
%FDA

%s4 ###
\section{Implications, Conclusions  and Suggestions}
\label{sec-conclude}

%s4.1 ###
\subsection{The Cost of Indiscrimination to Scientific~Research}

%My assertion is that both explanatory and predictive modeling are
%necessary for scientific development. Causal explanation leads to
%``understanding the world''. Explanatory modeling therefore assists in
%testing hypotheses. Prediction enables the assessment of the actual
%manifestation of causal mechanisms, the discovery of new relationships
%that can lead to new causal theory, and the evaluation of
%predictability. Given these roles of explanation and prediction, it is
%clear that explanatory and predictive modeling are complementary and
%both necessary for making scientific advances.

Currently, in many fields, statistical modeling is used nearly
exclusively for causal explanation. The consequence of neglecting to
include predictive modeling and testing alongside explanatory modeling
is losing the ability to test the relevance of existing theories and to
discover new causal mechanisms. Feelders (\citeyear{Feelders2002}) commented on the
field of economics: ``The pure hypothesis testing framework of economic
data analysis should be put aside to give more scope to learning from
the data. This closes the empirical cycle from observation to theory to
the testing of theories on new data.'' The current accelerated rate of
social, environmental, and technological changes creates a burning need
for new theories and for the examination of old theories in light of
the new realities.

A common practice due to the indiscrimination of explanation and
prediction is to erroneously infer predictive power from explanatory
power, which can lead to incorrect scientific and practical
conclusions. Colleagues from various fields confirmed this fact, and a
cursory search of their scientific literature brings up many examples.
For instance, in ecology an article intending to predict forest beetle
assemblages infers predictive power from explanatory power [``To study\ldots  predictive power$,
\ldots$\  we calculated the $R^2$''; ``We expect
predictabilities with $R^2$ of up to 0.6'' (Muller and Brandl, \citeyear{Muller2009})].
%; (2) fitting an explanatory model in conservation biology aimed to
%``predict those plant species likely to decline, persist or increase
%as a result of agricultural intensification'' while using inference to
%assess predictive performance (Afshartous and de Leeuw, \citeyear{Dorrough2008}); and (3) designing
%an experiment aimed at ``quantitative predictions of gene dispersal by
%pollen'', but using fitted values and goodness-of-fit to describe model
%``predictions'' (\citeauthor{Klein2006}, \citeyear{Klein2006}).
In economics, an article entitled ``The predictive power of zero
intelligence in financial markets'' (Farmer, Patelli and Zovko, \citeyear{Farmer2005}) infers predictive
power from a high $R^2$ value of a linear regression model.
In epidemiology, many studies rely on in-sample hazard ratios estimated
from Cox regression models to infer predictive power, reflecting an
indiscrimination between description and prediction. For instance, %
%Afshartous and de Leeuw (\citeyear{Huisman2007}) use them %``as indicators of the capacity of
%self-assessed health to predict mortality''. Similarly,
Nabi et al. (\citeyear{Nabi2010}) used hazard ratio estimates and statistical
significance ``to compare the predictive power of depression for
coronary heart disease with that of cerebrovascular disease.'' In
information systems, an article on ``Understanding
and predicting electronic commerce adoption'' (Pavlou and Fygenson, \citeyear{Pavlou2006})
incorrectly compared the predictive power of different models using
in-sample measures (``To examine the predictive power of the proposed
model, we compare it to four models in terms of $R^2$ adjusted''). These
examples are not singular, but rather they reflect the common
misunderstanding of predictive power in these and other fields.

%there is a stream of studies on ``the capacity of self-assessed health
%to predict mortality'', where in-sample hazard %ratios from a Cox
%regression model are ``used as indicators of the capacity of
%self-assessed health to predict %mortality'' (Afshartous and de Leeuw, \citeyear{Huisman2007}). In
%short, a conflation between explanation and prediction coupled with a
%monopoly of %explanatory modeling in a field can easily lead to
%erroneous conclusions.

Finally, a consequence of omitting predictive modeling from scientific
research is also a gap between research and practice. %Scientific
%research should benefit decision and policy making and daily life.
In an age where empirical research has become feasible in many fields,
the opportunity to bridge the gap between methodological development
and practical application can be easier to achieve through the
combination of explanatory and predictive modeling.

Finance is an example where practice is concerned with prediction
whereas academic research is focused on explaining. In particular,
there has been a reliance on a limited number of models that are
considered pillars of research, yet have proven to perform very poorly
in practice. For instance, the CAPM model and more recently the
Fama--French model are regression models that have been used for
explaining market behavior for the purpose of portfolio management, and
have been evaluated in terms
%. However, these models are evaluated based on their
of explanatory power (in-sample $R^2$ and residual analysis) and not
predictive accuracy.\footnote{Although in their paper Fama and French (\citeyear{Fama1993})
did split the sample into two parts, they did so for purposes of
testing the sensitivity of model estimates rather than for assessing
predictive accuracy.} More recently, researchers have begun recognizing
the distinction between in-sample explanatory power and out-of-sample
predictive power (Goyal and Welch, \citeyear{Goyal2007}), which has led to a discussion of
predictability magnitude and a search for predictively accurate
explanatory variables (Campbell and Thompson, \citeyear{Campbell2005}).
In terms of predictive modeling, the Chief Actuary of the Financial
Supervisory Authority of Sweden commented in 1999: ``there is a need
for models with predictive power for at least a very near future\ldots\
Given sufficient and relevant data this is an area for statistical
analysis, including cluster analysis and various kind of
structure-finding methods'' (Palmgren, \citeyear{Palmgren1999}). While there has been
some predictive modeling using
%Motivated by the ``long list of empirical features that traditional
%approaches have not been able to match'' %(Afshartous and de Leeuw, \citeyear{LeBaron2006}),
%predictive modeling has appeared in computational finance using
%algorithms such as
genetic algorithms (Chen, \citeyear{Chen2002}) and neural networks (Chakraborty and Sharma, \citeyear{Chakraborty2007}), it has been performed by practitioners and
nonfinance academic researchers and outside of the top academic journals.

%Similarly, in 2005 the Acting Commissioner of the Food and Drug
%Administration (FDA) in the US, recognizing the need for predictive
%modeling in the FDA's current explanatory modeling environment,
%commented:
%
% \begin{quote}
% Another aspect of the Critical Path initiative that applies to
%generic products is the search for tools, such as computer-based
%models, capable of predicting the success or failure of product
%development. We are planning to analyze data from a wide variety of
%drugs and formulations and from them develop predictive models for
%bioavailability and bioequivalence. (Afshartous and de Leeuw, \citeyear{Crawford2005})
% \end{quote}

In summary, the omission of predictive modeling for theory development
results not only in academic work becoming irrelevant to practice, but
also in creating a barrier to achieving significant scientific
progress, which is especially unfortunate as data become easier to
collect, store and access.

In the opposite direction, in fields that focus on predictive modeling,
the reason for omitting explanatory modeling must be sought. A
scientific field is usually defined by a cohesive body of theoretical
knowledge, which can be tested. Hence, some form of testing, whether
empirical or not, must be a component of the field. In areas such as
bioinformatics, where there is little theory and an abundance of data,
predictive models are pivotal in generating avenues for causal theory.
%%After a careful %step of constructing causal theories, explanatory
%modeling can be used for empirical testing, and then predictive
%%modeling can be used once again for assessing the relevance of the
%explanatory models.

%s4.2 ###
\subsection{Explanatory and Predictive Power: Two~Dimensions}
\label{subsec-2dim}

I have polarized explaining and predicting in this article in an effort
to highlight their fundamental differences. However, rather than
considering them as extremes on some continuum, I consider them as two
dimensions.\footnote{Similarly, descriptive models can be considered as
a third dimension, where yet different criteria are used for assessing
the strength of the descriptive model.}\tsup{,}\footnote{I thank Bill Langford
for the two-dimensional insight.} Explanatory power and predictive
accuracy are different qualities; a model will possess some level of each.

A related controversial question arises: must an explanatory model have
some level of predictive power to be considered scientifically useful?
And equally, must a predictive model have sufficient explanatory power
to be scientifically useful? For instance, some explanatory models that
cannot be tested for predictive accuracy yet constitute scientific
advances are Darwinian evolution theory and string theory in physics.
The latter produces currently untestable predictions
(Woit, \citeyear{Woit2006}, pages x--xii). Conversely, there exist predictive models that do
not properly ``explain'' yet are scientifically valuable. Galileo, in
his book \textit{Two New Sciences}, proposed a demonstration to
determine whether light was instantaneous. According to
Mackay and Oldford (\citeyear{Mackay2000}), Descartes gave the book a scathing review:

\begin{quote}

 The substantive criticisms are generally directed at
Galileo's not having identified the causes of the phenomena he
investigated. For most scientists at this time, and particularly for
Descartes, that is the whole point of science.

\end{quote}

\noindent
Similarly, consider predictive models that are based on a \textit
{wrong} explanation yet scientifically and practically they are
considered valuable. One well-known example is Ptolemaic astronomy,
which until recently was used for nautical navigation but is based on a
theory proven to be wrong long ago. While such examples are extreme, in
most cases models are likely to possess some level of both explanatory
and predictive power.

Considering predictive accuracy and explanatory power as two axes on a
two-dimensional plot would place different models ($f$), aimed either
at explanation or at prediction, on different areas of the plot. The
bi-dimensional approach implies that: (1) In terms of modeling, the
goal of a scientific study must be specified a priori in order to
optimize the criterion of interest; and (2) In terms of model
evaluation and scientific reporting, researchers should report \textit
{both the explanatory and predictive qualities} of their models. Even
if prediction is not the goal, the predictive qualities of a model
should be reported alongside its explanatory power so that it can be
fairly evaluated in terms of its capabilities and compared to other
models. Similarly, a predictive model might not require causal
explanation in order to be scientifically useful; however, reporting
its relation to causal theory is important for purposes of theory
building. The availability of information on a variety of predictive
and explanatory models along these two axes can shed light on both
predictive and causal aspects of scientific phenomena. The statistical
modeling process, as depicted in Figure \ref{fig-process}, should
include ``overall model performance'' in terms of both predictive and
explanatory qualities.

%s4.3 ###
\subsection{The Cost of Indiscrimination to the Field~of~Statistics}

Dissolving the ambiguity surrounding explanatory versus predictive
modeling is important for advancing our field itself. Recognizing that
statistical\break methodology has focused mainly on inference indicates an
important gap to be filled. %leaves much room for developing predictive
%methodology at various steps of the modeling %process.
While our literature contains predictive methodology for model
selection and predictive inference, there is scarce statistical
predictive methodology for other modeling steps, such as study design,
data collection, data preparation  and EDA, which present opportunities
for new research.
%Identifying such gaps are important for defining needed research
%areas.
Currently, the predictive void has been taken up the field of machine
learning and data mining. In fact, the differences, and some would say
rivalry, between the fields of statistics and data mining can be
attributed to their different goals of explaining versus predicting
even more than to factors such as data size. While statistical theory
has focused on model estimation, inference, and fit, machine learning
and data mining have concentrated on developing computationally
efficient predictive algorithms and tackling the bias--variance
trade-off in order to achieve high predictive accuracy.

Sharpening the distinction between explanatory and predictive modeling
can raise a new awareness of the strengths and limitations of existing
methods and practices, and might shed light on current controversies
within our field. One example is the disagreement in survey methodology
regarding the use of sampling weights in the analysis of survey data
(Little, \citeyear{Little2007}). Whereas some researchers advocate using weights to
reduce bias at the expense of increased variance, and others disagree,
might not the answer be related to the final goal?
%Perhaps when the %goal is inference %about the population mean, then
%weighting is beneficial; if the goal, however, is %predicting
%%individual observations, %then it may be better not to use weights,
%depending on the overall bias-variance %tradeoff.
%The reasoning behind such an argument is that weights can reduce bias
%but they also increase variance.

Another ambiguity that can benefit from an explanatory/predictive
distinction is the definition of parsimony. Some claim that predictive
models should be simpler than explanatory models: ``Simplicity is
relevant because complex families often do a bad job of predicting new
data, though they can be made to fit the old data quite well'' (Sober, \citeyear{Sober2002}). The same argument was given by Hastie, Tibshirani and Friedman (\citeyear{HTF2009}):
``Typically the more complex we make the model, the lower the bias but
the higher the variance.'' In contrast, some predictive models in
practice are very complex,\footnote{I thank Foster Provost from NYU for
this observation.} and indeed Breiman (\citeyear{Breiman2001}) commented: ``in some
cases predictive models are more complex in order to capture small
nuances that improve predictive accuracy.'' Zellner\break (\citeyear{Zellner2001}) used the
term ``sophisticatedly simple'' to define the quality of a ``good''
model. I would suggest that the definitions of parsimony and complexity
are task-dependent: predictive or explanatory. For example, an ``overly
complicated'' model in explanatory terms might prove ``sophisticatedly
simple'' for predictive purposes.

%s4.4 ###
\subsection{Closing Remarks and Suggestions}

The consequences from the explanatory/predictive distinction lead to
two proposed actions:
\begin{enumerate}[2.]
\item It is our responsibility to be aware of how statistical models
are used in research outside of statistics, why they are used in that
fashion, and in response to develop methods that support sound
scientific research. Such knowledge can be gained within our field by
inviting scientists from different disciplines to give talks at
statistics conferences and seminars, and to require graduate students
in statistics to read and present research papers from other disciplines.
\item As a discipline, we must acknowledge the difference between
explanatory, predictive  and descriptive modeling, and integrate it
into statistics education of statisticians and nonstatisticians, as
early as possible but most importantly in ``research methods'' courses.
This requires creating written materials that are easily accessible and
understandable by nonstatisticians. We should advocate both explanatory
and predictive modeling, clarify their differences and distinctive
scientific and practical uses, and disseminate tools and knowledge for
implementing both. One particular aspect to consider is advocating a
more careful use of terms such as ``predictors,'' ``predictions'' and
``predictive power,'' to reduce the effects of terminology on incorrect
scientific conclusions.
\end{enumerate}
Awareness of the distinction between explanatory and predictive
modeling, and of the different scientific functions that each serve, is
essential for the progress of scientific knowledge.

\begin{appendix}
\renewcommand{\theequation}{\arabic{equation}}
%s5 ###
\section*{Appendix: Is the ``True'' Model the Best Predictive Model? A Linear
Regression Example}
\label{sec-appendix}

\setcounter{equation}{1}

%In the following we show the breakdown of the prediction error into
%three terms: bias, variance,
%and unexplained variability. We use the following notation:
Consider ${\cal F}$ to be the true function relating constructs ${\cal
X}$ and ${\cal Y}$ and let us assume that $f$ is a valid
operationalization of ${\cal F}$. Choosing an intentionally biased
function $f^*$ in place of $f$ is clearly undesirable from a
theoretical--explanatory point of view. However, we will show that $f^*$
can be preferable to $f$ from a predictive standpoint.
%We use
%
%$Y =$ & Observed output\\
%$f(x) =$ & True model \\
%$f^*(x) =$ & Incorrect model specification \\
%$\hat{f}^*(x) =$ & Estimated model (based on incorrect model
%specification)

To illustrate this, consider the statistical model $f(x) = \beta_1
x_{1} + \beta_2 x_{2} + \epsilon$ which is assumed to be correctly
specified with respect to ${\cal F}$. Using data, we obtain the
estimated model $\hat{f}$, which has the properties
%
%e3 ###
%e2 ###
\begin{eqnarray}
\mathrm{Bias} &= &0, \\
\operatorname{Var}(\hat{f}(x)) &=& \operatorname{Var}(x_{1} \hat{\beta}_1 + x_{2} \hat
{\beta}_2)\nonumber
\\[-8pt]
\\[-8pt] &=& \sigma^2 x' (X'X)^{-1} x,\nonumber % = \sigma^2 H ,
\end{eqnarray}
where $x$ is the vector $x=[x_1, x_2]'$, and $X$ is the design matrix
based on both predictors. %, and $h_{i}$ is the $i$th diagonal element
%of the Hat matrix ($H=x(X'X)^{-1}x'$). %, based on using both
%predictors ($x=[X_1 , X_2]$).
Combining the squared bias with the variance gives
%
%e4 ###
\begin{eqnarray}
\mathrm{EPE} &=& E\bigl(Y-\hat{f}(x)\bigr)^2\nonumber\\
 &=& \sigma^2 + 0 + \sigma^2 x' (X'X)^{-1}
x\\
& = &\sigma^2\bigl(1+x' (X'X)^{-1} x\bigr).\nonumber
\end{eqnarray}

In comparison, consider the estimated underspecified form $\hat{f}^*(x)
= \hat{\gamma}_1 x_{1}$.
The bias and variance here are given by
Montgomery, Peck and Vining  (\citeyear{Montgomery2001}, pages~292--296):
\begin{eqnarray*}
\mathrm{Bias} &=& x_{1} \gamma_1 - (x_1 \beta_1 + x_{2} \beta_2)\\
& =& x_{1}
(x_1'x_1)^{-1} x_1'(x_1 \beta_1 +x_2 \beta_2)\\
&&{} - (x_1\beta_1 + x_2 \beta
_2), \\
%&=& x_1 (x_1' x_1)^{-1}x_1' x_2 \beta_2 - x_2 \beta_2 \\
\operatorname{Var}(\hat{f}^*(x)) &=& x_{1} \operatorname{Var}(\hat{\gamma}_1) x_1 =
\sigma^2 x_1(x_1^{\prime}x_1)^{-1}x_1.
\end{eqnarray*}
Combining the squared bias with the variance gives %leads to an MSE
%equal to
%
%e5 ###
\begin{eqnarray}
\mathrm{EPE} &=& %E(Y-\hat{f^*}(x))^2 =
\bigl( x_{1} (x_1' x_1)^{-1} x_1' x_2 \beta_2 - x_{2} \beta_2 \bigr)^2\nonumber
\\[-8pt]
\\[-8pt]&&{} + \sigma
^2\bigl(1+ x_1(x_1^{\prime}x_1)^{-1}x_1^{\prime}\bigr).\nonumber
\end{eqnarray}

Although the bias of the underspecified model $f^*(x)$ is larger than
that of $f(x)$, its variance can be
smaller, and in some cases so small that the overall EPE will be lower
for the underspecified
model. Wu, Harris and McAuley (\citeyear{Wu2007}) showed the general result for an underspecified
linear regression model with multiple
predictors. In particular, they showed that the underspecified model
that leaves out $q$ predictors
has a lower EPE when the following inequality holds:
%
%e6 ###
\begin{equation}
q \sigma^2 > \beta_2^{\prime} X_2^{\prime} ( I-H_1 ) X_2 \beta_2.
\end{equation}
This means that the underspecified model produces more accurate
predictions, in terms of lower
EPE, in the following situations:
\begin{itemize}
\item when the data are very noisy (large $\sigma$);
\item when the true absolute values of the left-out parameters (in our
example $\beta_2$) are small;
\item when the predictors are highly correlated; and
\item when the sample size is small or the range of left-out variables
is small.
\end{itemize}

The bottom line is nicely summarized by Hagerty and Srinivasan (\citeyear{Hagerty1991}): ``We note
that the practice in applied research of concluding that a model with a
higher predictive validity is ``truer,'' is not a valid inference. This
paper shows that a parsimonious but less true model can have a higher
predictive validity than a truer but less parsimonious model.''
\end{appendix}

\section*{Acknowledgments}
I thank two anonymous reviewers, the associate editor, and editor for
their suggestions and comments which improved this manuscript. I
express my gratitude to many colleagues for invaluable feedback and
fruitful discussion that have helped me develop the
explanatory/predictive argument presented in this article. I am
grateful to Otto Koppius (Erasmus) and Ravi Bapna (U Minnesota) for
familiarizing me with explanatory modeling in Information Systems, for
collaboratively pursuing prediction in this field, and for tireless
discussion of this work. I thank Ayala Cohen (Technion), Ralph Snyder
(Monash), Rob Hyndman (Monash)  and Bill Langford (RMIT) for detailed
feedback on earlier drafts of this article. Special thanks to Boaz
Shmueli and Raquelle Azran for their meticulous reading and discussions
of the manuscript. And special thanks for invaluable comments and
suggestions go to
Murray Aitkin (U Melbourne),
Yoav Benjamini (Tel\break Aviv~U),
Smarajit Bose (ISI),
Saibal Chattopadhyay (IIMC),
Ram Chellapah (Emory),
Etti Doveh (Technion),
Paul Feigin (Technion),
Paulo Goes\break (U~Arizona),
Avi Goldfarb (Toronto U),
Norma Hubele (ASU),
Ron Kenett (KPA Inc.),
Paul Lajbcygier (Monash),
Thomas Lumley (U Washington),
David Madigan (Columbia U),
Isaac Meilejson (Tel Aviv U),
Douglas Montgomery (ASU),
Amita Pal (ISI),
Don Poskitt (Monash),
Foster Provost (NYU),
Saharon Rosset (Tel Aviv U),
Jeffrey Simonoff (NYU)
and David Steinberg (Tel Aviv U).


\begin{thebibliography}{99}
%b1 ###
\bibitem[\protect\citeauthoryear{Afshartous and de Leeuw}{2005}]{Afshartous2005}
\textsc{Afshartous, D.} and \textsc{de Leeuw, J.}
(2005).
Prediction in multilevel models.
\textit{J. Educ.  Behav. Statist.}
\textbf{30} 109--139.

%b2 ###
\bibitem[\protect\citeauthoryear{Aitchison and Dunsmore}{1975}]{Aitchison1975a}
\textsc{Aitchison, J.} and \textsc{Dunsmore, I. R.}
(1975).
\textit{Statistical Prediction Analysis}.
Cambridge Univ. Press.
\MR{0408097}

%b3 ###
\bibitem[\protect\citeauthoryear{Bajari and Hortacsu}{2003}]{bajahort2003}
\textsc{Bajari, P.} and \textsc{Hortacsu, A.}
(2003).
The winner's curse, reserve prices and endogenous entry: Empirical
insights from ebay auctions.
\textit{Rand J. Econ.} \textbf{3} 329--355.

%b4 ###
\bibitem[\protect\citeauthoryear{Bajari and Hortacsu}{2004}]{bajahort2004}
\textsc{Bajari, P.} and \textsc{Hortacsu, A.}
(2004).
Economic insights from internet auctions.
\textit{J. Econ. Liter.} \textbf{42} 457--486.

%b5 ###
\bibitem[\protect\citeauthoryear{Bapna, Jank and Shmueli}{2008}]{BapnaJankShmueli2008}
\textsc{Bapna, R., Jank, W.} and \textsc{Shmueli, G.}
(2008).
Price formation and its dynamics in online auctions.
\textit{Decision Support Systems} \textbf{44} 641--656.

%b6 ###
\bibitem[\protect\citeauthoryear{Bell, Koren and Volinsky}{2008}]{Bell2008}
\textsc{Bell, R. M., Koren, Y.} and \textsc{Volinsky, C.}
(2008).
The {BellKor} 2008 solution to the {N}etflix {P}rize.

%b7 ###
\bibitem[\protect\citeauthoryear{Bell, Koren and Volinsky}{2010}]{Bell2010}
\textsc{Bell, R. M., Koren, Y.} and \textsc{Volinsky, C.}
(2010).
All together now: A perspective on the netflix prize.
\textit{Chance} \textbf{23} 24.

%b8 ###
\bibitem[\protect\citeauthoryear{Berk}{2008}]{Berk2008}
\textsc{Berk,} \textsc{R. A.}
(2008).
\textit{Statistical Learning from a Regression Perspective}.
Springer, New York.

%b9 ###
\bibitem[\protect\citeauthoryear{Bjornstad}{1990}]{Bjornstad1990}
\textsc{Bjornstad,} \textsc{J. F.}
(1990).
Predictive likelihood: A review.
\textit{Statist. Sci.} \textbf{5} 242--265.
\MR{1062578}

%b10 ###
\bibitem[\protect\citeauthoryear{Bohlmann and Hothorn}{2007}]{Bohlmann2007}
\textsc{Bohlmann, P.} and \textsc{Hothorn, T.}
(2007).
Boosting algorithms: Regularization, prediction and model fitting.
\textit{Statist. Sci.} \textbf{22} 477--505.
\MR{2420454}

%b11 ###
\bibitem[\protect\citeauthoryear{Breiman}{1996}]{Breiman1996}
\textsc{Breiman,} \textsc{L.}
(1996).
Bagging predictors.
\textit{Mach. Learn.} \textbf{24} 123--140.
\MR{1425957}

%b12 ###
\bibitem[\protect\citeauthoryear{Breiman}{2001a}]{Breiman2001a}
\textsc{Breiman,} \textsc{L.}
(2001a).
Random forests.
\textit{Mach. Learn.} \textbf{45} 5--32.

%b13 ###
\bibitem[\protect\citeauthoryear{Breiman}{2001b}]{Breiman2001}
\textsc{Breiman,} \textsc{L.}
(2001b).
Statistical modeling: The two cultures.
\textit{Statist. Sci.} \textbf{16} 199--215.
\MR{1874152}

%b14 ###
\bibitem[\protect\citeauthoryear{Brown, Vannucci and Fearn}{2002}]{Brown2002}
\textsc{Brown, P. J., Vannucci, M.} and \textsc{Fearn, T.}
(2002).
Bayes model averaging with selection of regressors.
\textit{J.  R. Stat. Soc. Ser.~B Stat. Methodol.} \textbf{64} 519--536.
\MR{1924304}

%b15 ###
\bibitem[\protect\citeauthoryear{Campbell and Thompson}{2005}]{Campbell2005}
\textsc{Campbell, J. Y.} and \textsc{Thompson, S. B.}
(2005).
Predicting excess stock returns out of sample: Can anything beat the
historical average?
Harvard Institute of Economic Research Working Paper 2084.

%b16 ###
\bibitem[\protect\citeauthoryear{Carte and Craig}{2003}]{Carte2003}
\textsc{Carte, T. A.} and \textsc{Craig, J. R.}
(2003).
In pursuit of moderation: Nine common errors and their solutions.
\textit{MIS Quart.} \textbf{27} 479--501.

%b17 ###
\bibitem[\protect\citeauthoryear{Chakraborty and Sharma}{2007}]{Chakraborty2007}
\textsc{Chakraborty, S.} and \textsc{Sharma, S. K.}
(2007).
Prediction of corporate financial health by artificial neural
network.
\textit{Int. J. Electron. Fin.} \textbf{1} 442--459.

%b18 ###
\bibitem[\protect\citeauthoryear{Chen}{2002}]{Chen2002}
\textsc{Chen,} \textsc{S.-H., Ed.}
(2002).
\textit{Genetic Algorithms and Genetic Programming in Computational
Finance}.
Kluwer, Dordrecht.

%b19 ###
\bibitem[\protect\citeauthoryear{Collopy, Adya  and Armstrong}{1994}]{Collopy1994}
\textsc{Collopy, F., Adya, M.} and \textsc{Armstrong, J.}
(1994).
Principles for examining predictive--validity---the case of
information-systems spending forecasts.
\textit{Inform. Syst. Res.} \textbf{5} 170--179.

%b20 ###
\bibitem[\protect\citeauthoryear{Dalkey and Helmer}{1963}]{Dalkey1963}
\textsc{Dalkey, N.} and \textsc{Helmer, O.}
(1963).
An experimental application of the delphi method to the use of
experts.
\textit{Manag. Sci.} \textbf{9} 458--467.

%b21 ###
\bibitem[\protect\citeauthoryear{Dawid}{1984}]{Dawid1984}
\textsc{Dawid,} \textsc{A. P.}
(1984).
Present position and potential developments: Some personal views:
Statistical theory: The prequential approach.
\textit{J. Roy. Statist. Soc. Ser.~A} \textbf{147} 278--292.
\MR{0763811}

%b22 ###
\bibitem[\protect\citeauthoryear{Ding and Simonoff}{2010}]{Ding2006}
\textsc{Ding, Y.} and \textsc{Simonoff, J.}
(2010).
An investigation of missing data methods for classification trees
applied to binary response data.
\textit{J. Mach. Learn. Res.} \textbf{11} 131--170.

%b23 ###
\bibitem[\protect\citeauthoryear{Domingos}{2000}]{Domingos2000}
\textsc{Domingos,} \textsc{P.}
(2000).
A unified bias--variance decomposition for zero--one and squared loss.
In \textit{Proceedings of the Seventeenth National Conference on
Artificial Intelligence}  564--569. AAAI Press, Austin, TX.

%b24 ###
\bibitem[\protect\citeauthoryear{Dowe, Gardner and Oppy}{2007}]{Dowe2007}
\textsc{Dowe, D. L., Gardner, S.} and \textsc{Oppy, G. R.}
(2007).
Bayes not bust! Why simplicity is no problem for Bayesians.
\textit{Br. J.   Philos. Sci.} \textbf{58} 709--754.
\MR{2375767}

%b25 ###
\bibitem[\protect\citeauthoryear{Dubin}{1969}]{Dubin1969}
\textsc{Dubin,} \textsc{R.}
(1969).
\textit{Theory Building}.
The Free Press, New York.

%b26 ###
\bibitem[\protect\citeauthoryear{Edwards and Bagozzi}{2000}]{Edwards2000}
\textsc{Edwards, J. R.} and \textsc{Bagozzi, R. P.}
(2000).
On the nature and direction of relationships between constructs.
\textit{Psychological Methods} \textbf{5} \textbf{2} 155--174.

%b27 ###
\bibitem[\protect\citeauthoryear{Ehrenberg and Bound}{1993}]{Ehrenberg1993}
\textsc{Ehrenberg, A.} and \textsc{Bound, J.}
(1993).
Predictability and prediction.
\textit{J. Roy. Statist. Soc. Ser.~A} \textbf{156} 167--206.

%b28 ###
\bibitem[\protect\citeauthoryear{Fama and French}{1993}]{Fama1993}
\textsc{Fama, E. F.} and \textsc{French, K. R.}
(1993).
Common risk factors in stock and bond returns.
\textit{J. Fin. Econ.} \textbf{33} 3--56.

%b29 ###
\bibitem[\protect\citeauthoryear{Farmer, Patelli and Zovko}{2005}]{Farmer2005}
\textsc{Farmer, J. D., Patelli, P.} and \textsc{Zovko, I. I. A. A.}
(2005).
The predictive power of zero intelligence in financial markets.
\textit{Proc. Natl. Acad.  Sci. USA} \textbf{102} 2254--2259.

%b30 ###
\bibitem[\protect\citeauthoryear{Fayyad, Grinstein and Wierse}{2002}]{Fayyad2002}
\textsc{Fayyad, U. M., Grinstein, G. G.} and \textsc{Wierse, A.}
(2002).
\textit{Information Visualization in Data Mining and Knowledge
Discovery}.
Morgan Kaufmann, San Francisco, CA.

%b31 ###
\bibitem[\protect\citeauthoryear{Feelders}{2002}]{Feelders2002}
\textsc{Feelders,} \textsc{A.}
(2002). Data mining in economic
science. In
\textit{Dealing with the Data Flood}      166--175.
STT/Beweton, Den Haag, The Netherlands.

%b32 ###
\bibitem[\protect\citeauthoryear{Findley and Parzen}{1998}]{Findley1998}
\textsc{Findley, D. Y.} and \textsc{Parzen, E.}
(1998). A conversation
with Hirotsugo Akaike. In
\textit{Selected Papers of Hirotugu Akaike}    3--16.
Springer, New York.
\MR{1486823}

%b33 ###
\bibitem[\protect\citeauthoryear{Forster}{2002}]{Forster2002}
\textsc{Forster,} \textsc{M.}
(2002).
Predictive accuracy as an achievable goal of science.
\textit{Philos. Sci.} \textbf{69} S124--S134.

%b34 ###
\bibitem[\protect\citeauthoryear{Forster and Sober}{1994}]{Forster1994}
\textsc{Forster, M.} and \textsc{Sober, E.}
(1994).
How to tell when simpler, more unified, or less ad-hoc theories will
provide more accurate predictions.
\textit{Br. J.   Philos.   Sci.} \textbf{45} 1--35.
\MR{1277464}

%b35 ###
\bibitem[\protect\citeauthoryear{Friedman}{1997}]{Friedman1997}
\textsc{Friedman,} \textsc{J. H.}
(1997).
On bias, variance, 0$/$1-loss, and the curse-of-dimensionality.
\textit{Data Mining and Knowledge Discovery} \textbf{1} 55--77.

%b36 ###
\bibitem[\protect\citeauthoryear{Gefen, Karahanna and Straub}{2003}]{Gefen2003}
\textsc{Gefen, D., Karahanna, E.} and \textsc{Straub, D.}
(2003).
Trust and {TAM} in online shopping: An integrated model.
\textit{MIS Quart.} \textbf{27} 51--90.

%b37 ###
\bibitem[\protect\citeauthoryear{Geisser}{1975}]{Geisser1975}
\textsc{Geisser,} \textsc{S.}
(1975).
The predictive sample reuse method with applications.
\textit{J. Amer. Statist. Assoc.} \textbf{70} 320--328.

%b38 ###
\bibitem[\protect\citeauthoryear{Geisser}{1993}]{Geisser1993}
\textsc{Geisser,} \textsc{S.}
(1993).
\textit{Predictive Inference: An Introduction}.
Chapman and Hall, London.
\MR{1252174}

%b39 ###
\bibitem[\protect\citeauthoryear{Gelman et al.}{2003}]{Gelman2003}
\textsc{Gelman, A., Carlin, J. B., Stern, H. S.} and \textsc{Rubin, D. B.}
(2003).
\textit{Bayesian Data Analysis}, 2nd ed.
Chapman \& Hall/CRC  New York/Boca Raton, FL.
\MR{1385925}

%b40 ###
\bibitem[\protect\citeauthoryear{Ghani and Simmons}{2004}]{Ghani2004}
\textsc{Ghani, R.} and \textsc{Simmons, H.}
(2004).
Predicting the end-price of online auctions.
In \textit{International Workshop on Data Mining and Adaptive Modelling
Methods for Economics and Management}, Pisa, Italy.

%b41 ###
\bibitem[\protect\citeauthoryear{Goyal and Welch}{2007}]{Goyal2007}
\textsc{Goyal, A.} and \textsc{Welch, I.}
(2007).
A comprehensive look at the empirical performance of equity premium
prediction.
\textit{Rev. Fin. Stud.} \textbf{21} 1455--1508.

%b42 ###
\bibitem[\protect\citeauthoryear{Granger}{1969}]{Granger1969}
\textsc{Granger,} \textsc{C.}
(1969).
Investigating causal relations by econometric models and
cross-spectral methods.
\textit{Econometrica} \textbf{37} 424--438.

%b43 ###
\bibitem[\protect\citeauthoryear{Greenberg and Parks}{1997}]{Greenberg1997}
\textsc{Greenberg, E.} and \textsc{Parks, R. P.}
(1997).
A predictive approach to model selection and multicollinearity.
\textit{J. Appl. Econom.} \textbf{12} 67--75.

%b44 ###
\bibitem[\protect\citeauthoryear{Gurbaxani and Mendelson}{1990}]{Gurbaxani1990}
\textsc{Gurbaxani, V.} and \textsc{Mendelson, H.}
(1990).
An integrative model of information systems spending growth.
\textit{Inform. Syst. Res.} \textbf{1} 23--46.

%b45 ###
\bibitem[\protect\citeauthoryear{Gurbaxani and Mendelson}{1994}]{Gurbaxani1994}
\textsc{Gurbaxani, V.} and \textsc{Mendelson, H.}
(1994).
Modeling vs. forecasting---the case of information-systems spending.
\textit{Inform. Syst. Res.} \textbf{5} 180--190.

%b46 ###
\bibitem[\protect\citeauthoryear{Hagerty and Srinivasan}{1991}]{Hagerty1991}
\textsc{Hagerty, M. R.} and \textsc{Srinivasan, S.}
(1991).
Comparing the predictive powers of alternative multiple regression
models.
\textit{Psychometrika} \textbf{56} 77--85.
\MR{1115296}

%b47 ###
\bibitem[\protect\citeauthoryear{Hastie, Tibshirani and Friedman}{2009}]{HTF2009}
\textsc{Hastie, T., Tibshirani, R.} and \textsc{Friedman, J. H.}
(2009).
\textit{The Elements of Statistical Learning: Data Mining, Inference,
and Prediction}, 2nd ed.
Springer, New York.
\MR{1851606}

%b48 ###
\bibitem[\protect\citeauthoryear{Hausman}{1978}]{Hausman1978}
\textsc{Hausman,} \textsc{J. A.}
(1978).
Specification tests in econometrics.
\textit{Econometrica} \textbf{46} 1251--1271.
\MR{0513692}

%b49 ###
\bibitem[\protect\citeauthoryear{Helmer and Rescher}{1959}]{Helmer1959}
\textsc{Helmer, O.} and \textsc{Rescher, N.}
(1959).
On the epistemology of the inexact sciences.
\textit{Manag. Sci.} \textbf{5} 25--52.

%b50 ###
\bibitem[\protect\citeauthoryear{Hempel and Oppenheim}{1948}]{Hempel1948}
\textsc{Hempel, C.} and \textsc{Oppenheim, P.}
(1948).
Studies in the logic of explanation.
\textit{Philos. Sci.} \textbf{15} 135--175.

%b51 ###
\bibitem[\protect\citeauthoryear{Hitchcock and Sober}{2004}]{Hitchcock2004}
\textsc{Hitchcock, C.} and \textsc{Sober, E.}
(2004).
Prediction versus accommodation and the risk of overfitting.
\textit{Br. J. Philos.   Sci.} \textbf{55} 1--34.

%b52 ###
\bibitem[\protect\citeauthoryear{Jaccard}{2001}]{Jaccard2001}
\textsc{Jaccard,} \textsc{J.}
(2001).
\textit{Interaction Effects in Logistic Regression}.
SAGE Publications, Thousand Oaks, CA.

%b53 ###
\bibitem[\protect\citeauthoryear{Jank and Shmueli}{2010}]{JankShmueli2010}
\textsc{Jank, W.} and \textsc{Shmueli, G.}
(2010).
\textit{Modeling Online Auctions}.
Wiley, New York.

%b54 ###
\bibitem[\protect\citeauthoryear{Jank, Shmueli and Wang}{2008}]{Jank2008}
\textsc{Jank, W., Shmueli, G.} and \textsc{Wang, S.}
(2008). Modeling
price dynamics in online auctions via regression trees.
In
\textit{Statistical Methods in eCommerce Research}.
Wiley, New York.
\MR{2414052}

%b55 ###
\bibitem[\protect\citeauthoryear{Jap and Naik}{2008}]{Jap2008}
\textsc{Jap, S.} and \textsc{Naik, P.}
(2008).
Bidanalyzer: A method for estimation and selection of dynamic bidding
models.
\textit{Marketing Sci.} \textbf{27} 949--960.

%b56 ###
\bibitem[\protect\citeauthoryear{Johnson and Geisser}{1983}]{Johnsone1983}
\textsc{Johnson, W.} and \textsc{Geisser, S.}
(1983).
A predictive view of the detection and characterization of
influential observations in regression analysis.
\textit{J. Amer. Statist. Assoc.} \textbf{78} 137--144.
\MR{0696858}

%b57 ###
\bibitem[\protect\citeauthoryear{Kadane and Lazar}{2004}]{Kadane2004}
\textsc{Kadane, J. B.} and \textsc{Lazar, N. A.}
(2004).
Methods and criteria for model selection.
\textit{J. Amer. Statist. Soc.} \textbf{99} 279--290.
\MR{2061890}

%b58 ###
\bibitem[\protect\citeauthoryear{Kendall and Stuart}{1977}]{Kendall1977}
\textsc{Kendall, M.} and \textsc{Stuart, A.}
(1977).
\textit{The Advanced Theory of Statistics}    \textbf{1}, 4th ed.
  Griffin, London.

%b59 ###
\bibitem[\protect\citeauthoryear{Konishi and Kitagawa}{2007}]{Konishi2007}
\textsc{Konishi, S.} and \textsc{Kitagawa, G.}
(2007).
\textit{Information Criteria and Statistical Modeling}.
Springer, New York.
\MR{2367855}

%b60 ###
\bibitem[\protect\citeauthoryear{Krishna}{2002}]{kris2002}
\textsc{Krishna,} \textsc{V.}
(2002).
\textit{Auction Theory}.
Academic Press, San Diego, CA.

%b61 ###
\bibitem[\protect\citeauthoryear{Little}{2007}]{Little2007}
\textsc{Little,} \textsc{R. J. A.}
(2007).
Should we use the survey weights to weight?
JPSM Distinguished Lecture, Univ. Maryland.

%b62 ###
\bibitem[\protect\citeauthoryear{Little and Rubin}{2002}]{Little2002}
\textsc{Little, R. J. A.} and \textsc{Rubin, D. B.}
(2002).
\textit{Statistical Analysis with Missing Data}.
Wiley, New York.
\MR{1925014}

%b63 ###
\bibitem[\protect\citeauthoryear{Lucking-Reiley et al.}{2007}]{Reiley2007}
\textsc{Lucking-Reiley, D., Bryan, D., Prasad, N.} and \textsc{Reeves,~D.}
(2007).
Pennies from ebay: The determinants of price in online auctions.
\textit{J. Indust. Econ.} \textbf{55} 223--233.

%b64 ###
\bibitem[\protect\citeauthoryear{Mackay and Oldford}{2000}]{Mackay2000}
\textsc{Mackay, R. J.} and \textsc{Oldford, R. W.}
(2000).
Scientific method, statistical method, and the speed of light.
Working  Paper 2000-02, Dept.  Statistics and Actuarial
Science, Univ. Waterloo.
\MR{1847825}

%b65 ###
\bibitem[\protect\citeauthoryear{Makridakis, Wheelwright and Hyndman}{1998}]{Makridakis1997}
\textsc{Makridakis, S. G., Wheelwright, S. C.} and \textsc{Hyndman, R.~J.}
(1998).
\textit{Forecasting: Methods and Applications}, 3rd ed.
Wiley, New York.

%b66 ###
\bibitem[\protect\citeauthoryear{Montgomery, Peck and Vining}{2001}]{Montgomery2001}
\textsc{Montgomery, D., Peck, E. A.} and \textsc{Vining, G. G.}
(2001).
\textit{Introduction to Linear Regression Analysis}.
Wiley, New York.
\MR{1820113}

%b67 ###
\bibitem[\protect\citeauthoryear{Mosteller and Tukey}{1977}]{MostellerTukey1977}
\textsc{Mosteller, F.} and \textsc{Tukey, J. W.}
(1977).
\textit{Data Analysis and Regression}.
Addison-Wesley, Reading, MA.

%b68 ###
\bibitem[\protect\citeauthoryear{Muller and Brandl}{2009}]{Muller2009}
\textsc{Muller, J.} and \textsc{Brandl, R.}
(2009).
Assessing biodiversity by remote sensing in mountainous terrain: The
potential of lidar to predict forest beetle assemblages.
\textit{J. Appl. Ecol.} \textbf{46} 897--905.

%b69 ###
\bibitem[\protect\citeauthoryear{Nabi et al.}{2010}]{Nabi2010}
\textsc{Nabi, J., Kivim\"{a}ki, M., Suominen, S., Koskenvuo, M.} and \textsc{Vahtera, J.}
(2010).
Does depression predict coronary heart diseaseand cerebrovascular
disease equally well? The health and social support prospective cohort study.
\textit{Int. J. Epidemiol.}  \textbf{39} 1016--1024.

%b70 ###
\bibitem[\protect\citeauthoryear{Palmgren}{1999}]{Palmgren1999}
\textsc{Palmgren,} \textsc{B.}
(1999).
The need for financial models.
 \textit{ERCIM News}  \textbf{38} 8--9.

%b71 ###
\bibitem[\protect\citeauthoryear{Parzen}{2001}]{Parzen2001}
\textsc{Parzen,} \textsc{E.}
(2001).
Comment on statistical modeling: The two cultures.
\textit{Statist. Sci.} \textbf{16} 224--226.
\MR{1874152}

%b72 ###
\bibitem[\protect\citeauthoryear{Patzer}{1995}]{Patzer1995}
\textsc{Patzer,} \textsc{G. L.}
(1995).
\textit{Using Secondary Data in Marketing Research: United States and
Worldwide}.
Greenwood Publishing, Westport, CT.

%b73 ###
\bibitem[\protect\citeauthoryear{Pavlou and Fygenson}{2006}]{Pavlou2006}
\textsc{Pavlou, P.} and \textsc{Fygenson, M.}
(2006).
Understanding and predicting electronic commerce adoption: An
extension of the theory of planned behavior.
\textit{Mis Quart.} \textbf{30} 115--143.

%b74 ###
\bibitem[\protect\citeauthoryear{Pearl}{1995}]{Pearl1995}
\textsc{Pearl,} \textsc{J.}
(1995).
Causal diagrams for empirical research.
\textit{Biometrika} \textbf{82} 669--709.
\MR{1380809}

%b75 ###
\bibitem[\protect\citeauthoryear{Rosenbaum and Rubin}{1983}]{Rosenbaum1983}
\textsc{Rosenbaum, P.} and \textsc{Rubin, D. B.}
(1983).
The central role of the propensity score in observational studies for
causal effects.
\textit{Biometrika} \textbf{70} 41--55.
\MR{0742974}

%b76 ###
\bibitem[\protect\citeauthoryear{Rubin}{1997}]{Rubin1997}
\textsc{Rubin,} \textsc{D. B.}
(1997).
Estimating causal effects from large data sets using propensity
scores.
\textit{Ann. Intern. Med.}  \textbf{127}   757--763.

%b77 ###
\bibitem[\protect\citeauthoryear{Saar-Tsechansky and Provost}{2007}]{Saar-Tsechansky2007}
\textsc{Saar-Tsechansky, M.} and \textsc{Provost, F.}
(2007).
Handling missing features when applying classification models.
\textit{J. Mach. Learn. Res.} \textbf{8} 1625--1657.

%b78 ###
\bibitem[\protect\citeauthoryear{Sarle}{1998}]{Sarle1998}
\textsc{Sarle,} \textsc{W. S.}
(1998).
Prediction with missing inputs.
In  \textit{JCIS 98 Proceedings} (P. Wang,  ed.)    \textbf{II}
399--402. Research Triangle Park, Durham, NC.

%b79 ###
\bibitem[\protect\citeauthoryear{Seni and Elder}{2010}]{seni2010}
\textsc{Seni, G.} and \textsc{Elder, J. F.}
(2010).
\textit{Ensemble Methods in Data Mining: Improving Accuracy Through
Combining Predictions (Synthesis Lectures on Data Mining and Knowledge
Discovery)}.
Morgan and Claypool, San Rafael, CA.

%b80 ###
\bibitem[\protect\citeauthoryear{Shafer}{1996}]{Shafer1996}
\textsc{Shafer,} \textsc{G.}
(1996).
\textit{The Art of Causal Conjecture}.
MIT Press, Cambridge, MA.

%b81 ###
\bibitem[\protect\citeauthoryear{Schapire}{1999}]{Shapire1999}
\textsc{Schapire,} \textsc{R. E.}
(1999).
A brief introduction to boosting.
In \textit{Proceedings of the Sixth International Joint Conference on
Artificial Intelligence} 1401--1406. Stockholm, Sweden.

%b82 ###
\bibitem[\protect\citeauthoryear{Shmueli and Koppius}{2010}]{ShmueliKoppius2010}
\textsc{Shmueli, G.} and \textsc{Koppius, O. R.}
(2010).
Predictive analytics in information systems research.
\textit{MIS Quart.}  To appear.

%b83 ###
\bibitem[\protect\citeauthoryear{Simon}{2001}]{Simon2001}
\textsc{Simon,} \textsc{H. A.}
(2001). Science seeks parsimony, not simplicity: Searching for
pattern in phenomena.
In \textit{Simplicity, Inference and Modelling: Keeping it Sophisticatedly
Simple}     32--72.
Cambridge Univ. Press.
\MR{1932928}

%b84 ###
\bibitem[\protect\citeauthoryear{Sober}{2002}]{Sober2002}
\textsc{Sober,} \textsc{E.}
(2002).
Instrumentalism, parsimony, and the Akaike framework.
\textit{Philos. Sci.} \textbf{69} S112--S123.

%b85 ###
\bibitem[\protect\citeauthoryear{Song and Witt}{2000}]{Song2000}
\textsc{Song, H.} and \textsc{Witt, S. F.}
(2000).
\textit{Tourism Demand Modelling and Forecasting: Modern Econometric
Approaches}. Pergamon Press,
Oxford.

%b86 ###
\bibitem[\protect\citeauthoryear{Spirtes, Glymour and Scheines}{2000}]{Spirtes2000}
\textsc{Spirtes, P., Glymour, C.} and \textsc{Scheines, R.}
(2000).
\textit{Causation, Prediction, and Search}, 2nd ed.
MIT Press, Cambridge, MA.
\MR{1815675}

%b87 ###
\bibitem[\protect\citeauthoryear{Stone}{1974}]{Stone1974}
\textsc{Stone,} \textsc{M.}
(1974).
Cross-validatory choice and assesment of statistical predictions
(with discussion).
\textit{J. Roy. Statist. Soc. Ser.~B} \textbf{39} 111--147.
\MR{0356377}

%b88 ###
\bibitem[\protect\citeauthoryear{Taleb}{2007}]{Taleb2007}
\textsc{Taleb,} \textsc{N.}
(2007).
\textit{The Black Swan}.
Penguin Books, London.

%b89 ###
\bibitem[\protect\citeauthoryear{Van Maanen, Sorensen and Mitchell}{2007}]{Van2007}
\textsc{Van Maanen, J., Sorensen, J.} and \textsc{Mitchell, T.}
(2007).
The interplay between theory and method.
\textit{Acad. Manag. Rev.} \textbf{32} 1145--1154.

%b90 ###
\bibitem[\protect\citeauthoryear{Vaughan and Berry}{2005}]{Vaughan2005}
\textsc{Vaughan, T. S.} and \textsc{Berry, K. E.}
(2005).
Using Monte Carlo techniques to demonstrate the meaning and
implications of multicollinearity.
\textit{J. Statist. Educ.} \textbf{13} online.

%b91 ###
\bibitem[\protect\citeauthoryear{Wallis}{1980}]{Wallis1980}
\textsc{Wallis,} \textsc{W. A.}
(1980).
The statistical research group, 1942--1945.
\textit{J. Amer. Statist. Assoc.} \textbf{75} 320--330.
\MR{0577363}

%b92 ###
\bibitem[\protect\citeauthoryear{Wang, Jank and Shmueli}{2008}]{WangJankShmueli2008}
\textsc{Wang, S., Jank, W.} and \textsc{Shmueli, G.}
(2008).
Explaining and forecasting online auction prices and their dynamics
using functional data analysis.
\textit{J. Business   Econ. Statist.} \textbf{26} 144--160.\
\MR{2420144}

%b93 ###
\bibitem[\protect\citeauthoryear{Winkelmann}{2008}]{Winkelmann2008}
\textsc{Winkelmann,} \textsc{R.}
(2008).
\textit{Econometric Analysis of Count Data}, 5th ed.
Springer, New York.
\MR{2148271}

%b94 ###
\bibitem[\protect\citeauthoryear{Woit}{2006}]{Woit2006}
\textsc{Woit,} \textsc{P.}
(2006).
\textit{Not Even Wrong: The Failure of String Theory and the Search for
Unity in Physical Law}. Jonathan Cope, London.
\MR{2245858}

%b95 ###
\bibitem[\protect\citeauthoryear{Wu, Harris and McAuley}{2007}]{Wu2007}
\textsc{Wu, S., Harris, T.} and \textsc{McAuley, K.}
(2007).
The use of simplified or misspecified models: Linear case.
\textit{Canad. J. Chem. Eng.} \textbf{85} 386--398.

%b96 ###
\bibitem[\protect\citeauthoryear{Zellner}{1962}]{Zellner1962}
\textsc{Zellner,} \textsc{A.}
(1962).
An efficient method of estimating seemingly unrelated regression
equations and tests for aggregation bias.
\textit{J.~Amer. Statist. Assoc.} \textbf{57} 348--368.
\MR{0139235}

%b97 ###
\bibitem[\protect\citeauthoryear{Zellner}{2001}]{Zellner2001}
\textsc{Zellner,} \textsc{A.}
(2001).
     Keep it sophisticatedly simple. In \textit{Simplicity, Inference and Modelling: Keeping It Sophisticatedly
Simple}  242--261.
Cambridge Univ. Press.
\MR{1932939}

%b98 ###
\bibitem[\protect\citeauthoryear{Zhang, Jank and Shmueli}{2010}]{Zhang2009}
\textsc{Zhang, S., Jank, W.} and \textsc{Shmueli, G.}
(2010).
Real-time forecasting of online auctions via functional $k$-nearest
neighbors.
\textit{Int. J. Forecast.}  \textbf{26}
666--683.

\end{thebibliography}
\end{document}